\documentstyle[aps,twocolumn]{revtex}

\begin{document}
\author{A. V. Mahajan$^1$\cite{byline}, H. Alloul$^1$, G. Collin$^2$, and J. F.
Marucco$^3$}
\title{$^{89}$Y NMR Probe of Zn Induced Local Magnetism in YBa$_2$(Cu$_{1-y}$Zn$_{y}
$)$_3$O$_{6+x}$}
\address{$^1$Laboratoire de Physique des Solides, URA 2\\
CNRS,\\
Univ. Paris-Sud, 91405 Orsay, France\\
$^2$LLB, CE Saclay, CEA-CNRS, 91191, Gif Sur Yvette,\\
France\\
$^3$Laboratoire des Compos\'{e}s Non-Stoechiom\'{e}triques,\\
Universit\'{e} Paris-Sud, 91405, Orsay, France}
\date{\today}


\twocolumn[\hsize\textwidth\columnwidth\hsize\csname@twocolumnfalse\endcsname
\maketitle

\begin{abstract}
We present detailed data and analysis of the effects of Zn substitution on
the planar Cu site in YBa$_2$Cu$_3$O$_{6+x}$ (YBCO$_{6+x}$) as evidenced
from our $^{89}$Y NMR measurements on oriented powders. For $x<<1$ we find
additional NMR lines which are associated with the Zn substitution. From our
data on the intensities and temperature dependence of the shift, width, and
spin-lattice relaxation rate of these resonances, we conclude that the
spinless Zn 3$d$$^{10}$ state induces local moments on the near-neighbour ($%
nn$) Cu atoms. Additionally, we conjecture that the local moments actually
extend to the farther Cu atoms with the magnetization alternating in sign at
subsequent $nn$ sites. We show that this analysis is compatible with ESR
data taken on dilute Gd doped (on the Y site) and on neutron scattering data
reported recently on Zn substituted YBCO$_{6 + x}$. For optimally doped
compounds $^{89}$Y $nn$ resonances are not detected,  but a large $T$%
-dependent contribution to the $^{89}$Y NMR linewidth is evidenced  and is
also attributed to the occurence of a weak induced local moment near the Zn.
These results are compatible with macroscopic magnetic measurements
performed on YBCO$_{6 + x}$ samples prepared specifically in order to
minimize the content of impurity phases.  We find significant differences
between the present results on the underdoped YBCO$_{6 + x}$ samples and $%
^{27}$Al NMR data taken on Al$^{3+}$ substituted on the Cu site in optimally
doped La$_2$CuO$_4$. Further experimental work is needed to clarify the
detailed evolution of the impurity induced magnetism with hole content in
the cuprates.
\end{abstract}

]
\section{Introduction}

It is now experimentally well established that the CuO$_{2}$ planes are
responsible for the magnetic and superconducting properties of the cuprates.
However the interconnection between these two properties is still an
essential but unanswered question. Understanding the normal state of the
cuprates is still a prerequisite for any theoretical approach to the
microscopic origin of High Temperature Superconductivity. In the recent
past, considerable interest has been aroused due to the detection of a
pseudo-gap in the spin excitation spectrum of the cuprates for underdoped
materials (the word pseudo is prefixed because although a strong decrease in
the intensity of excited states is detected well above the superconducting
transition temperature $T_{c}$, a real gap is only detected below $T_{c}$).
The first indications of a pseudo-gap were provided in the microscopic NMR
measurements of the susceptibility of the CuO$_{2}$ planes. $^{89}$Y NMR
shifts in YBCO$_{6+x}$ of Alloul {\it et al.} \cite{alloulohno} were found
to decrease markedly with decreasing $T$. The large decrease of the static
susceptibility was interpreted to be due to an opening of a pseudo-gap in
the homogeneous ${\bf q}=0$ excitations of the system. In the underdoped ($x$
$<$ 1 for YBCO$_{6+x}$) high-T$_{c}$ cuprates, a similar decrease of the
spin-lattice relaxation rate $^{63}$Cu $1/T_{1}T$ \cite{warren}, which is
dominated by the imaginary part of the susceptibility at the AF wave vector $%
{\bf q}=(\pi, \pi )$ was also indicative of a pseudo-gap in the spin
excitations at this wave vector \cite{berthier}.

The inelastic neutron scattering experiments which followed \cite
{rossat,ginsberg}, clearly confirmed the existence of this pseudo spin-gap
at $(\pi ,\pi )$, and allowed measurements of the frequency dependence of
the excitations. The temperatures at which these two pseudo-gaps begin to
open are found to be different, and it is not clear at present whether they
signal different cross-overs between distinct states or a single cross-over
phenomenon. The latter case would imply a wave-vector dependence of the
pseudo-gap. Presently, the existence of a pseudogap for the underdoped high-$%
T_{c}$ superconductors has been detected by many techniques such as
transport, photoemission, etc. While various explanations are proposed for
the pseudo-gaps, it is believed that the essential physics of the normal
state (and perhaps the superconducting state), of the cuprates might be
linked to it.

In order to better characterise the properties of the cuprates it has
appeared quite important to understand their modifications due to impurities
or disorder. Atomic substitutions on the planar Cu site are naturally found
to be the most detrimental to superconductivity, while modification in the
charge reservoir chains mainly yield changes in hole doping. For such
studies the YBCO$_{6+x}$ system is particularly suitable, as the variation
of impurity induced magnetism with hole doping can be studied by merely
changing the oxygen content $x$ for a given impurity content. In classical
superconductors, $T_c$ is mainly affected by magnetic impurity
substitutions. In cuprates it has been shown that even a non-magnetic
impurity like Zn ($3d^{10}$), which substitutes on the Cu site of the CuO$_2$
plane strongly decreases the superconducting transition temperature $T_c$
(about 10.6 K/\% Zn for $x$ = 1). It has also been anticipated \cite{fink}
and then shown experimentally that although Zn itself is non-magnetic, it
induces a modification of the magnetic properties of the correlated spin
system of the CuO$_2$ planes \cite{alloul2}. Using $^{89}$Y NMR we have
further shown, in the preliminary report of the present work \cite{mahajan},
that local magnetic moments are induced on the $nn$ Cu of the Zn substituent
in the CuO$_2$ plane. Two important result have been demonstrated:

i) the $q = 0$ pseudo-gap was found unaffected by Zn even when $T_c$ is
reduced to zero for YBCO$_{6.6}$,

ii) The magnitude of the induced local moment is strongly dependent on the
carrier concentration \cite{mendels1}.

Since our reports, other experimental evidence by NMR in YBCO \cite
{dupree1,walstedtdupree}, in YBa$_2$Cu$_4$O$_8$ (1248) \cite{williams1}, in
La$_2$CuO$_4$ \cite{ishida2}, or ESR in YBCO and 1248 \cite{janossy}, have
confirmed that the occurrence of local moment induced by non-magnetic
impurities on the Cu sites is a general property of cuprates. The local
moments have been observed as well in macroscopic bulk susceptibility data 
\cite{mendels1,cooper,mendels,jps1}. The Zn-induced modifications of the
magnetic excitations both in the superconducting and the normal state have
been studied by neutron scattering \cite{kakurai,sidis}. Also, electrical
transport \cite{ong,mizuhashi}, and thermal properties \cite{loram} of
substituted high-$T_c$ cuprates have been investigated.

However some studies have concluded that the susceptibility near the Zn does
not exhibit a Curie behaviour, at least for $x$ = 1, or that the AF
correlations were destroyed in the vicinity of the Zn substituents \cite
{janossy,ishida1}. Also some data have been interpreted as due to the total
disappearance of the pseudo-gap in the vicinity of Zn. Finally, the dynamics
of the local moment \cite{ishida2} appears to be quite different in La$_2$CuO%
$_4$:Al than in our results.

To clarify the situation, we present in this article an extended report of
our experimental data, and perform an exhaustive comparison with the
literature. We examine in detail the effect of Zn on $^{89}$Y NMR, in
oriented powders of YBCO$_{6+x}$:Zn$_y$ with 0.5\% $\leq$ $y$ $\leq$ 4 \%,
for $x$ = 0.64 and 1. NMR, being a local probe, provides useful information
about the impurity induced short-range and long-range effects in the metal
via an analysis of the lineshape, Knight shift, and linewidth. In section
II, we present the experimental details regarding sample preparation and the
procedures adopted for NMR measurements. In section III, a thorough
description of the results of our NMR work, allows us to highlight the
differences in the effect of Zn doping in underdoped and overdoped YBCO. In
section IV, the NMR shift, linewidth, and relaxation rate data are analyzed
considering Zn induced local moments on the neighbouring Cu. A contrasting
comparison with the other experimental results introduced in this section is
contained in the last subsection of IV. In the conclusion section we
summarize our overall view of the experimental situation, and discuss
several theoretical works which have paid some attention to the induced
magnetism in cuprates.

\section{Experimental Details}

Samples of YBCO$_{6+x}$:Zn were prepared by conventional solid state
reaction techniques as described elsewhere \cite{alloul3}. Large (single
crystal) grain (size $>$ 50 microns) samples were made which were finely
ground before oxygenation (see Ref.~\cite{laurence} for further details
regarding characterisation). In order to prepare the samples with maximum
oxygen content, oxygenation was done at $\sim$ 300 $^{\circ}$C for a long
period ($>$ 10 days) which ensured homogeneity of oxygen content. For
preparing samples with a reduced oxygen content, the maximally oxidized
samples were treated in vacuum in a thermobalance at variable temperatures,
up to 450 $^{\circ}$C. The samples were quenched to room temperature when
the equilibrium oxygen content was reached.

In the case of YBCO$_{6+x}$ (without Zn), when the maximum oxidized samples
were deoxidized to a point where the sample decomposed, the weight loss
corresponded to $\delta$x (= x$_{max}$- x$_{min}$) = 1.0. However, on
addition of Zn, the actual maximum value of $\delta$x which could be
reached, progressively decreases and equals 0.92 for 4\% Zn. Since for the
Zn doped samples, the ortho-tetra structural phase transition still takes
place for an oxygen content of x$_{min}$ + 0.45 (as in YBCO$_{6+x}$ without
Zn) \cite{mendels2}, it appears that x$_{min}$ = 6.0 in the Zn doped samples
while x$_{max}$ linearly decreases from 7.0 for 0\% Zn to 6.92 for 4\% Zn.
These samples with specific oxygen and zinc contents were then fixed in
Stycast 1266 and cured overnight in a field of 7.5 Tesla in order to orient
the grains with the $c$ axis aligned along the applied field direction.

NMR measurements were performed by standard pulsed NMR techniques. We
observed the spin echos after a $\pi/2-\pi$ sequence followed by a $%
3\pi/2-\pi$ sequence. A perfect inversion of the spin-echo in the latter
relative to the first sequence ensured the correctness of the $\pi/2$ pulse
length (about 13 $\mu$sec at room temperature). The $^{89}$Y shift was
measured with respect to a standard YCl$_3$ solution.

The $^{89}$Y spin-lattice relaxation time $T_1$ was determined using a $%
\pi/2-\pi$ sequence, with a repetition time $t_{rep}$. An exponential fit of
the nuclear magnetization (obtained from a Fourier Transform, FT, of the
time domain spin-echo signal) as a function of $t_{rep}$ allowed us to
deduce $T_1$ .

For YBCO$_{6.64}$:Zn, $nn$ resonances are seen (see Fig.~1) at low
temperatures (T $<$ 150 K). The relative intensity of these $nn$ resonances
could be enhanced by repeating the pulse sequence at a fast rate ($t_{rep}$ $%
\sim$ 20 sec). Indeed, the $T_1$ of the outermost satellite ($\sim$ 10 sec),
was found smaller than that of the main line ($\sim$ 100 sec).The fact that
the latter has a reduced intensity in such an experimental condition thus
allows us to fix accurately the position and the width of the $nn$
resonances. The mainline being the narrowest and the most intense, its
position and width were easily determined with a long repetition time,
allowing full recovery of the mainline signal. Using the positions and
widths of the $nn$ resonances determined in the manner indicated above, the
relative intensities of the various lines in the spectrum were determined
(by fitting the lineshape to a sum of three gaussians), for $t_{rep}$ $>$ 5$%
T_1$ of the slowest recovering component, so that all the components had
fully recovered. The $T_1$'s of the individual lines were determined from an
exponential fit of their intensities (in the FT) with respect to the
repetition time. Each $T_1$ measurement took about 15 hours.

\section{Results}

In the following, we present the doping and temperature dependence of
various NMR parameters in YBCO$_{6+x}$:Zn. We shall first report the
existence of additional NMR lines detected in the underdoped samples. Their
characteristics (shift, width, and intensity) enable us to associate them
with $^{89}$Y nuclei near-neighbours of the Zn substituted on the CuO$_2$
planes (Section III-A-1). The results on the main resonance line, which
corresponds to $^{89}$Y sites far from the substituted Zn, are reported next
and compared to those in the pure system (Section III-A-2). Spin-lattice
relaxation data on the $nn$ and main resonance lines are reported in Section
III-B.

\subsection{Resonance line shift and width}

\subsubsection{Near neighbour resonances}

We will argue here that the additional resonances detected in YBCO$_{6.64}$%
:Zn are not seen in YBCO$_7$:Zn and that the additional resonances are
intrinsic and a direct effect of Zn substitution.

As seen in Fig.~1 the $nn$-resonance positions depend on the sample
orientation with respect to the applied field. Furthermore (see Fig.~2), the
relative intensity of the outer line increases with Zn content while its
position is unchanged. We also see that, YBCO$_7$:Zn spectra (Fig.~3 (a)) do
not show additional lines in the temperature range of our measurements (80 $%
< $ $T$ $<$ 350). While this is unambiguously evident for the outermost
resonance, the absence of the middle resonance in spectra of YBCO$_7$:Zn is
perhaps not immediately obvious. By measuring the YBCO$_7$:Zn lineshape with
a fast repetition rate (so that the middle resonance might be enhanced,
relative to the mainline, due to its shorter $T_1$), we see (Fig.~3 (b)) in
fact, that the lineshape of YBCO$_7$:Zn is unaltered by fast repetition (up
to one-fourth of $T_1$ of the mainline). If there is any change, it is in
fact the high frequency tail that has a somewhat reduced relative intensity,
indicating that the tail has a longer $T_1$. This is in keeping with our
understanding that the upper tail in the lineshape of YBCO$_7$:Zn appears
due to those regions of the sample which are not fully oxidized and hence
have a longer $T_1$. In short, YBCO$_{6.64}$:Zn has additional $^{89}$Y
resonances while YBCO$_7$:Zn does not. In view of the abovementioned facts,
the additional resonances are not due to spurious phases since those should
be present independent of the oxygen content (the deoxygenated samples are
obtained merely by vacuum reduction of YBCO$_7$:Zn at low $T$ ($< $ 450 $%
^{\circ}$C)).

In order to identify the origin of these lines we have therefore performed
quantitative analyses of the spectral intensity. Experimental lineshapes for
YBCO$_{6.64}$:Zn, obtained with repetition times much longer than the
spin-lattice relaxation times $T_1$, were fitted to a sum of three
gaussians, where the line position and width of the two outer lines had been
reliably fixed from the short repetition time spectra. The spectra along
with the fits are shown in Fig.~4, while the variation of their intensity as
a function of Zn content is shown in Fig.~5.

It is of course quite natural to expect that the most affected outer line
should be associated with the Y nuclei $nn$ to the Zn atoms. In the dilute
limit, the intensity from a purely statistical occupancy of a single
neighbouring site of Y by Zn, for an in plane concentration $c$, is 8$%
c(1-c)^{7}$ for the 1$^{st}$ shell (curve A in Fig.~5).  But, as Zn induces
a significant shift of the 2$^{nd}$ $nn$ Y sites as well, we also need to
ensure that the 2$^{nd}$ $nn$ to Y is unoccupied by Zn. The corresponding
intensity for a purely random statistical occupancy would then be modified
to 8$c(1-c)^{15}$ (curve B in Fig.~5) which yields a smaller intensity for
large Zn concentrations. We see that the intensity of the outer line is then
consistent with that of the 1$^{st}$ $nn$ shell, {\it assuming that all the
Zn are substituted in the planes} ( in Fig.~5 we have taken $c = 1.5y$). As
for the middle resonance, the expression for the intensity due to the
occupancy of a single Y 2$^{nd}$ $nn$ site by Zn (with the 1$^{st}$ $nn$
unoccupied) is 16$c(1-c)^{23}$ in the dilute limit, which is much smaller
than the experimental intensity. If the 2$^{nd}$ and 3$^{rd}$ $nn$ are
occupied by Zn with the 1$^{st}$ $nn$ unoccupied, the intensity would be (8$%
c(1-c)^{7}$ + 16$c(1-c)^{15}$)(1-c)$^{8}$ (curve C in Fig.~5). The
assignment for the middle resonance is not so clear, but for dilute samples
its intensity is consistent with that of total occupancy of the 2$^{nd}$ and
3$^{rd}$ $nn$, with the 1$^{st}$ $nn$ unoccupied by Zn.

The $T$-dependence of the $nn$ line-shifts shown in Fig.~6 is seen to be
Curie-like with a negative hyperfine coupling. This Curie-like behaviour is
usually observed for local moments and justifies this denomination that we
introduced in \cite{alloul2}, although the actual magnitude and exact origin
of this local moment behaviour will only become clear hereafter.

The linewidth of the $nn$ resonances is found to increase with decreasing $T$
(Fig.~7). The linewidth which increases with Zn concentration may be
associated with the RKKY-like interaction between the Zn induced local
moments. With an increase in the concentration of local moments, we might
expect a frozen magnetic state (most probably a disordered spin-glass) at
some point in temperature. An estimate for this is provided by a simple
analysis in Section IV-E.

\subsubsection{Main resonance}

The temperature dependence of the shifts, $\Delta K(T)$, of the mainline for
YBCO$_{6.64}$:Zn is shown in Fig.~8. As reported before \cite{alloul2}, the
mainline shift does not significantly depend on the Zn content, and the
average carrier density at long distance from Zn carrier density is
therefore nearly unaffected by Zn substitution, at least for dilute
concentrations of Zn for which the sample is still metallic. A slight offset
with respect to the pure YBCO$_{6.64}$ is however evident at higher Zn
doping levels and might be due to a small increase in the carrier
concentration. Similar slight offsets are detected for the $^{17}$O NMR
shift of these compounds \cite{bobroff}, but would rather correspond to a
minute decrease in the hole content. In this latter case, a slightly
incomplete oxygen loading of the starting samples might result from the fact
that it has to be achieved in sealed vials, and not in a flowing oxygen
atmosphere, to facilitate $^{17}$O enrichment.

It should be mentioned that in Ref.~\cite{alloul2}, the mainline shift in an
unoriented YBCO$_{6.64}$:Zn$_{4\%}$ sample, had showed a slight upturn at
low temperatures. Since the satellite intensities constitute a significant
fraction here and cannot be clearly distinguished from the mainline, the
line-position obtained from the peak did not represent the true position of
the main line. In the present work, the different components of the spectra
have been analysed in the fits with different repetition times so that the
true position of the main resonance is deduced and shows no upturn.

The width of the mainline for YBCO$_{6.64}$:Zn$_{y}$ (see Fig.~9) has a $T$%
-dependence similar to that of pure YBCO$_{6.64}$, in that it initially
decreases with decreasing temperature and then shows an increase below 120 K
which is sample dependent. In the pure system, the linewidth can only be
associated with a small macroscopic distribution of chain oxygen content
(which we estimate of about $\pm$0.02 for most oxygen contents) which
results in a distribution of shifts at high temperatures. The $T$%
-dependencies of the shifts around the YBCO$_{6.64}$ composition are such
that while the shifts are measurably different at room temperature, their
magnitudes become nearly the same at low-$T$ \cite{alloulohno} and the NMR
line becomes narrower. Therefore the width due to a distribution of oxygen
content decreases at low-$T$. The magnitude of the width increases with Zn
doping, partly due to long-distance effects of the spin-polarisation from
the Zn-induced local moments. The $T$-dependence of the $^{17}$O NMR width
is found to be much larger and provides supplementary information which is
analysed in detail by Bobroff {\it et al.} \cite{bobroff}.

Turning to the fully oxygenated YBCO$_7$:Zn samples, we find that here again
the mainline shift is nearly independent of Zn concentration (Fig.~10).
However, due to the broadening of the line at low-temperatures, we cannot
unambiguously determine whether the maximum seen in the $T$ dependence of $%
^{89}$Y NMR shift of pure YBCO$_7$, with T(K$_{max}$) only slightly larger
than $T_c$ (and having a possible connection to the pseudo-gap) has shifted
to lower temperatures or altogether disappeared. The Curie-like broadening
(Fig.~11) in YBCO$_7$:Zn is indicative of a distribution of magnetic
contributions to the line positions. This RKKY-like broadening must
originate from a magnetic state which develops around the doped Zn.

For oxygen contents intermediate between O$_{7}$ and O$_{6.6}$, the $T$%
-dependence of the $^{89}$Y shift in YBCO$_{6+x}$:Zn is qualitatively
similar to that in YBCO$_{6+x}$ (Fig.~12(a)). However, the sharp decrease in
the shift that occurs around 100 K for the slightly oxygen depleted samples
(YBCO$_{6.95}$ or so) is absent in the Zn doped samples where the decrease
in the shift is more gradual with $T$. This might again be due to the
difficulty in defining accurately the oxygen content (and therefore the hole
concentration for the Zn substituted samples and especially for the large 4
\% Zn concentration which has been systematically investigated. As for the
outer resonance, we did not perform systematic investigations versus oxygen
content. It is however clear in the spectra of Fig.~12(b), that the
low-frequency tail which monitors the position of this outer resonance is
progressively nearer to the central line when the oxygen content is
increased. Further, this outer resonance even disappears for $x = 0.92$
which corresponds to the maximum oxygen content for 4 \% Zn. The NMR shift
of the outer resonance with respect to the mainline is therefore
progressively reduced with increasing hole content.

\subsection{Spin-lattice relaxation}

Next, we present the results of spin-lattice relaxation measurements for
YBCO $_{6.64}$:Zn. The data were obtained on the $nn$ lines and the mainline
as detailed in the previous section. Representative spectra for various
values of $t_{rep}$ are shown in Fig.~13. The resulting magnetisation
recovery for the three lines for the data of Fig.~13 are shown in Fig.~14.
Exponential fits have been found to apply in all cases as illustrated in
Fig.~14.

Taking such data is obviously not straightforward. A high enough signal to
noise ratio is required, as seen for our spectra displayed in Fig.~13. Other
publications on $^{89}$Y NMR in YBa$_{2}$Cu$_{4}$O$_{8}$:Zn \cite
{williams1,williams2} are completely bereft of $T_{1}$ data, which
corroborates the difficulty in obtaining good data. The relaxation rate of
the additional lines (other than the mainline) is seen (Fig.~15) to be
strongly enhanced resulting from local moment fluctuations as is discussed
in section IV-A2. The outermost satellite is the most affected which
indicates that it must result from having Zn as its 1$^{st}nn$. The mainline 
$T_{1}$ is nearly unaffected which shows that, for dilute concentrations of
Zn, the planar dynamic susceptibility far from Zn is unaffected, in
accordance with the NMR shift data.

In YBCO$_7$, Dupree {\it et al.} \cite{dupree1} measured the effect of Zn
doping on $T_1$ at room temperature. They found that the $^{89}$Y
spin-lattice relaxation rate was strongly enhanced on Zn substitution.
However, our data were in complete disagreement with theirs. We therefore
repeated measurements on various batches of samples and at various
temperatures. In all cases we found that the nuclear magnetisation recovery
fits well to a single exponential (Fig.~16) and that the resulting $T_1$ and
its $T$-dependence is not significantly different from that of pure YBCO$_7$%
, as can be seen in Fig.~17. We must therefore conclude that limited
accuracy was responsible for the observation done by Dupree {\it et al.} 
\cite{dupree1}. As there are no discernible resonances in addition to the
main line in these experiments on YBCO$_7$:Zn, the implication is a much
weaker induced moment in YBCO$_7$, compared to YBCO$_{6.64}$:Zn, in
agreement with our bulk susceptibility data \cite{mendels1,mendels}.

\section{Analysis of the experimental results}

\subsection{Local moments in YBCO$_{6.64}$:Zn}

\subsubsection{$nn$ NMR shifts}

We recall here, that the distinct, well defined resonances that we have
observed in YBCO$_{6.64}$ correspond to Y near neighbour sites of the
substituted Zn. The Curie-like $T$-dependence of the position of the first
near-neighbour line, and the shortening of its T$_1$ at low-$T$ are striking
experimental evidence of the occurrence of Zn induced local moments. The
location, spatial extent and dynamics of these moments in YBCO$_{6.64}$ will
be discussed first. Occurrence of local moments for the slightly overdoped
composition YBCO$_{7}$ is also established through the induced long-distance
perturbation of the host-spin-magnetization.

The Zn induced local moments are quite clearly located in the vicinity of
the Zn, and dominantly on the four nearest neighbour O or Cu orbitals. In
what follows, we shall perform extensive comparisons of the $^{89}$Y NMR
shift with the Zn induced Curie contribution to the spin susceptibility
(expressed per mole Zn) \cite{mendels1,mendels},

\begin{equation}
\chi_c = \frac{C_M}{T} = \frac{N_A p^2_{eff}}{3k_BT}
\end{equation}

Here $N_A$ is the avogadro number and $p_{eff}$ is the effective moment.
These comparisons will allow us first to rule out a localisation of the
moments on the O orbitals and then to demonstrate that the local moment is
distributed on the Cu orbitals. Furthermore, assuming that the transferred
hyperfine couplings are not modified by Zn substitution, we will show that
our analysis is consistent with a locally AF state extended over a few
lattice sites.

A local moment could also be present around Zn if a hole were trapped on the
near-neighbour oxygen orbitals. Two quite different physical situations
would occur, depending on whether the local moment is located on the $%
p_{\pi} $ or $p_{\sigma}$ orbitals. Since the $p_{\pi}$ are directly admixed
with the Y $s$ orbitals, a strong positive hyperfine coupling would result,
contrary to our observation of a negative Curie contribution to the shift.
Therefore, the present experiment implies that this shift component can only
be induced through the oxygen $p_{\sigma}$ orbitals.

In undoped YBCO$_{6+x}$, the Y NMR shift arises from a coupling of the Y
nuclear spin with the small fraction of holes on the O(2$p_{\sigma}$)
orbitals due to their {\it covalency} with the Cu(3$d_{x^2-y^2}$) holes,
while the spin-polarization of the doped holes themselves is negligible \cite
{alloulohno}. This has been deduced from the fact that the covalent
admixture of the O(2$p_{\sigma}$) orbital with the Cu(3$d_{x^2-y^2}$)
orbital is about 10 \% \cite{hybrid}, which implies that the hyperfine
coupling to the oxygen holes should be 10 times larger than its coupling to
the Cu(3$d_{x^2-y^2}$) holes.

Let us first consider the possibility that the Curie contribution comes from
holes localised on the four O($2p_{\sigma}$) orbitals near Zn. The 1$^{st}$ $%
nn$ Y site has six O $nn$ which we assume are nearly unaffected by Zn and
two O $nn$ which would exhibit Zn induced Curie magnetism. The net $^{89}$Y
shift would be written as 
\begin{equation}
\Delta K_{1}^{\alpha}=(6/8)K_{s}^{\alpha}+K_{c}^{\alpha}+\delta_{ 1}^{
\alpha}
\end{equation}
where index $\alpha$ refers to a principal direction, $K_{s}^{\alpha}$ is
the spin shift of the mainline, $K_{c}^{\alpha}$ =2$C_{s}^{\alpha}/T$ is the
Curie contribution to the spin shift due to its two 1$^{st}$ $nn$ O with a
moment, and $\delta_{1}^{\alpha}$ is the chemical shift. A least- squares
fit to the data in Fig.~18(a) allows us to extract the two unknown
parameters $C_{s}^{\alpha}$ and $\delta_{1}^{\alpha}$. The chemical shift
values thus obtained are $\delta_{1}^c$ $(\delta_{1}^{ab})$ =144 (163) $\pm$
10 ppm. The values of $C_{s}^{\alpha}$ are found to be -14000 (-12300) $\pm$
500 ppm K for H$\mid \mid$c (H$\mid \mid$ab).

Let us compare then these shifts with the macroscopic susceptibility $\chi_c$%
, with $\mu_B K_{c}$ = $2H_{hf}\chi_c/4$, if the moment is distributed on
the four O(2$p_{\sigma}$) near neighbour orbitals to the Zn. Using $C_M$ =
9.2 $\times$ 10$^{-2}$ emu K/mole Zn \cite{mendels1,mendels} and the Curie
term in the shift deduced above, we get H$_{hf}$ = -1.6 kG. This is of the
order of the hyperfine coupling expected with Cu and nearly 10 times smaller
than that expected with oxygen. Moreover, we point out that a 2$^{nd}$ $nn$
Y to Zn would not be coupled to the moment on the oxygen (a local moment on
oxygen is unlikely to be spread over more than 4 sites since it would
presumably arise from hole localisation). Hence, a second line in addition
to the main line should not be observed, contrary to the data from our
experiment. One could imagine that one has both, localised hole and weakly
affected $nn$ Cu. But this would require a large susceptibility on the Cu to
give the strong shift of the 2$^{nd}$ $nn$ line compared to the mainline.
This eliminates the oxygen $p_{\sigma}$ as a possible site for local moments.

If, however, the satellite shift is modelled as coming from a hyperfine
coupling to the local moments residing on the Cu $d_{x^{2}- y^{2}}$
orbitals, the relevant equation for the 1$^{st}$ $nn$ shift in our model is 
\begin{equation}
\Delta K_{1}^{\alpha }=(5/8)K_{s}^{\alpha }+K_{c}^{\alpha }+\delta
_{1}^{\alpha }
\end{equation}
>From a fit of the data to this equation, the chemical shift values obtained
(see Fig.~18(b)), $\delta _{1}^{c}$ $(\delta _{1}^{ab})$ =100 (140) $\pm $
10 ppm, are only slightly different from $\delta _{1}^{c}$ $(\delta
_{1}^{ab})$ =165 (150) ppm found in the pure material \cite{alloul2}. The
values of $C_{s}^{\alpha }$ are found to be -13100 (-11600) $\pm $ 500 ppm.
This implies a hyperfine field $H_{hf}$ $\approx $ -3.2 kG/Cu which is
slightly larger than that for the pure material ($\approx $ -2 kG). Such a
modification of $H_{hf}$ could be attributed to a corresponding change of
the Cu(3$d_{x^{2}-y^{2}}$)-O(2$p_{\sigma }$) hybridisation due to a
displacement of the 1$^{st}$ $nn$ oxygen to Zn . Alternatively, if the
hyperfine coupling stays unchanged, the actual susceptibility on the 1$^{st}$
$nn$ Cu, $\chi (1)$, is larger than $\chi _{c}/4$. This would imply that
further copper ions would have a magnetization anti-parallel to the applied
field, which might be expected if the local moment develops as an AF
correlated cloud of copper lattice sites. Considering this possibility, the
Cu 2$^{nd}$ $nn$ to Zn will bear a small negative susceptibility $\chi (2)$.

Fig.~19 illustrates schematically the location and the orientation of the Zn
induced local moments. In such a model, the shifts of the 1$^{st}$ and the 2$%
^{nd}$ $nn$ Y will be as follows; 
\begin{equation}
\Delta K_{1}^{\alpha }=(4/8)K_{s}^{\alpha }+2K_{1}^{\alpha }+K_{2}^{\alpha
}+\delta _{1}^{\alpha }
\end{equation}
\begin{equation}
\Delta K_{2}^{\alpha }=(6/8)K_{s}^{\alpha }+K_{1}^{\alpha }+K_{2}^{\alpha
}+\delta _{2}^{\alpha }
\end{equation}
Here we can differentiate the hyperfine fields for the first and second $nn$
using $\mu _{B}K_{1}^{\alpha }=H_{hf}(1)\chi (1)$ and $\mu _{B}K_{2}^{\alpha
}=H_{hf}(2)\chi (2)$. We can then fit $\Delta K_{1}^{\alpha
}-(4/8)K_{s}^{\alpha }$ to a Curie term in addition to a constant and
likewise for $\Delta K_{2}^{\alpha }-(6/8)K_{s}^{\alpha }$. A fit of the
observed shifts of the 1$^{st}$ and the 2$^{nd}$ $nn$ Y to these equations
yields the following values for the corresponding Curie terms; $C_{1}^{c}$ =
-14,530 ppm K, $C_{2}^{c}$ = + 4630 ppm K. The corresponding chemical shifts
for the 1$^{st}$ and the 2$^{nd}$ $nn$ Y are found to be 75 and 144 ppm,
respectively. Assuming the same hyperfine coupling for 1$^{st}$ $nn$ and 2$%
^{nd}$ $nn$ Y, $\chi (2)=-\chi (1)/3$ and therefore the macroscopic
susceptibility $\chi _{c}=8\chi (1)/3$. Using $\mu_{B}K_{1}^{%
\alpha}=H_{hf}(1)\chi (1)$, we get a hyperfine field of about -2.35 kG which
is closer to the value of -2 kG in the undoped compound. Although this
picture is then compatible with the experimental results, the accuracy of
the data is not sufficient to ascertain its validity. The fact that the
intensity of the middle resonance cannot be assigned solely to the 2$^{nd}$ $%
nn$ of Zn implies that further near neighbour sites of the Zn should be
taken into account. More accuracy would also be required to take into
account the 3$^{rd}$ $nn$ and try to estimate the size of the AF correlated
region around the Zn site.

\subsubsection{Spin-lattice relaxation}

We next turn to a discussion of the spin-lattice relaxation rate which is
expressed as, 
\begin{equation}
\frac{1}{T_1T} \propto \Sigma _{{\bf q}} A^2({\bf q}) \frac{\chi^{\prime
\prime}({\bf q}, \omega)}{\omega}
\end{equation}
where $A({\bf q})$ is the coupling to the magnetic fluctuations at wave
vector ${\bf q}$ \cite{moriya}. The O and Y nuclei are at symmetry positions
with respect to Cu so that $A({\bf q}_{AF})=0$, and the fluctuations at $%
{\bf q}_{AF}$ are filtered at these two sites. Consequently, in YBCO$_{6+x}$%
, the $T$-dependence of $(T_1T)^{-1}$ for Y and O is different from that of
Cu which is dominated by the fluctuations at ${\bf q}_{AF}$ \cite{ginsberg}.
On adding Zn, the symmetry around the 1$^{st}$ $nn$ Y is broken and this Y
site becomes then sensitive to the magnetic fluctuations on the neighbouring
copper ions (either intrinsic to the pure compound or due to the local
moment). Similarly, the magnetic fluctuations are no longer symmetric on the
2$^{nd}$ $nn$, so that the enhanced relaxation rate at this site is also
connected to local moment fluctuations.

The sharp increase of $(T_{1}T)^{-1}$ at low-$T$ on the Y $nn$ nuclei (much
faster than the corresponding variation on the $^{63}$Cu nuclei in the pure
compounds) is a direct proof that the local moment fluctuations are {\it not}
those of the Cu hole spins of the pure compound. The very existence of a
Curie contribution to the spin susceptibility indeed clearly points out that
the fluctuation of the Cu hole spins in the vicinity of the Zn are much
slower than those of the pure host. The present data for $(T_{1}T)^{-1}$ on
the near-neighbour nuclei are then good proof of the slow fluctuations of
the local moment.

In the case of local moments in noble metal hosts, the spin-lattice
relaxation of host nuclei nearby the local moment is totally dominated at
low-$T$ by the fluctuations of the local moment (the usual Korringa process
via conduction electrons is somewhat smaller) \cite{alloul1}. The situation
here is quite similar for the Y 1$^{st}$ $nn$ of the Zn. We can therefore
consider that this nuclear spin is coupled to the 2 $nn$ coppers which bear
susceptibilities $\chi _{c}({\bf q},\omega )$ if we neglect the contribution
to $T_{1}$ of the uncompensated Cu spin 2$^{nd}$ $nn$ to Zn (Fig.~19). The
relaxation rate at low-$T$ is then given by, 
\begin{equation}
\frac{1}{T_{1}}=\frac{2k_{B}T}{\hbar ^{2}}(\frac{\gamma _{n}}{\gamma _{e}}
)^{2}H_{hf}(1)^{2}\Sigma (\frac{\chi _{c}^{\prime \prime }({\bf q},\omega )}{%
\omega})
\end{equation}
where $\gamma _{n}/\gamma _{e}$ is the ratio of the nuclear and the
electronic gyromagnetic ratios and $H_{hf}(1)$ is the hyperfine coupling. In
the limit $\omega \rightarrow 0$, the summation is given by $\chi
_{c}(T)\tau /2\pi $ where $\chi _{c}(T)$ is the local moment susceptibility
and $\tau $ is the relaxation time of the local moment spin. The fluctuation
rate of the local moment spin is usually made up of two contributions 
\begin{equation}
\frac{1}{\tau }=\frac{1}{\tau _{ex}}+\frac{1}{\tau _{int}}
\end{equation}
where the first term corresponds to the single Zn impurity local moment
relaxation to the host spin bath, for instance through the exchange with the
conduction electron spins and the second would correspond to fluctuations
due the coupling between the Zn induced local moments which depends on Zn
concentration.

For instance, for dilute local moments in noble metal hosts

\begin{equation}
\frac{1}{\tau _{ex}}=(\frac{4\pi }{\hbar })(k_{B}T)(J_{ex}\rho (\epsilon
_{F}))^{2}
\end{equation}
if a Korringa relation holds ($J_{ex}$ is the coupling of the local moments
to the band). In that case, the second term is $\tau _{int}^{- 1}=\omega
_{int}/2\pi $ with $\omega _{int}^{2}=8J_{int}^{2}zS(S+1)/3\hbar ^{2}$ where 
$z\propto c$ is the number of nearest neigbour spins and $J_{int}$ is the
conduction electron mediated coupling between impurity spins.

In our case, the $T_{1}$ values for the near neighbour resonances did not
depend markedly on the Zn content, so that the single impurity induced
relaxation $1/\tau _{ex}$ dominates the results. Further, the local moment
spin susceptibility is Curie-like down to low temperatures. A small
Curie-Weiss correction $\chi =C/(T+\theta )$ with $\theta $ $\simeq 4$ K is
observed for YBCO$_{6.64}$:Zn$_{4\%}$ \cite{mendels1}. This is in agreement
with the observed spin freezing temperature of about 3 K in the sample \cite
{mendels1}. All these results therefore allow to conclude consistently that
the spin-lattice relaxation is dominated by the spin fluctuations of the
isolated Zn induced local moment. This should then be written as 
\begin{equation}
\frac{1}{T_{1}}=\frac{2k_{B}T}{\hbar ^{2}}(\frac{\gamma _{n}}{\gamma _{e}}
)^{2}H_{hf}(1)^{2}\frac{\chi _{c}(T)\tau _{ex}}{2\pi }
\end{equation}
Since $\chi _{c}T$ is constant, $1/T_{1}$ is proportional to $\tau _{ex}$.
In the conventional metallic case, where $\rho (\epsilon _{F})$ is
independent of the temperature, the spin-lattice relaxation rate $1/T_{1}$
of the host nuclei (near impurities) is observed to follow a Curie- like law 
$1/T_{1}$ $\propto $ $C/T$ \cite{alloul1}. The present case is clearly more
complicated since the host metal itself is strongly correlated, which
results in the pseudo-gap of the static spin susceptibility and in the $%
1/(T_{1}T)$ behaviour for the $^{63}$Cu. As the local moment positions are
commensurate with the Cu hole spin system, we might anticipate that a
similar anomaly might occur for $1/(\tau _{ex}T)$, which scales with $%
T_{1}/T $. We have therefore plotted in Fig.~20 this quantity for the three
data points for Y near neighbours to Zn. Although the data are clearly
insufficient, they suggest a maximum for $1/(\tau _{ex}T)$, quite analogous
to that seen for $^{63}$Cu $T_{1}$.

\subsection{Comparison with other experiments}

Following our early report \cite{mahajan}, some related experiments have
been performed on the impurity substituted cuprates. In various cases, the
authors have drawn conclusions which are not in complete agreement with our
results and sometimes disagree totally. These contradictions are therefore
considered in the following.

\subsubsection{Gd ESR}

Janossy {\it et al.} \cite{janossy} have done Gd$^{3+}$ ESR, in 1\% Gd
substituted YBCO$_{6+x}$, which are in principle quite similar to our
experiments. Indeed, the Gd electronic moment and the $^{89}$Y nuclear spin
are coupled to the copper hole spins by similar transferred couplings. They
find that the $g$-shift of the Gd ESR has the same $T$ dependence as the $%
^{89}$Y NMR shift. They have therefore used the Gd probe in Y$_{0.99}$Gd $%
_{0.01}$BaCuO$_{6 + x}$:Zn samples as well. The Gd spectrum should result
from a simple scaling with the $^{89}$Y NMR spectrum through the ratio of
the hyperfine couplings, as the relative positions of the main and satellite
lines scale as the ratio of $(\nu H_{hf})$ where $\nu$ is the operating
frequency and 
\begin{equation}
(\delta \nu/\nu) = H_{hf}\chi
\end{equation}
The fact that they did not detect any $nn$ resonances might cast some doubt
on the meaning of our results. However, we insist here that one should also
consider scaling of the relaxation rates to arrive at reliable conclusions.
The relaxation rates scale as 
\begin{equation}
1/T_1 \propto (\gamma H_{hf})^2
\end{equation}
and therefore contribute quite differently to the broadening of the spectra.
In the case of NMR, the linewidth $\Delta \nu$ is governed by the
susceptibility distribution which scales as $\delta \nu$, while in ESR, the $%
T_1$ process is so efficient that it contributes significantly to (and even
dominates) the ESR linewidth. Using our detailed data for $^{89}$Y $nn$ NMR,
we can easily calculate the expected contributions for the Gd $nn$ ESR,
through Eq. (11)-(12). Here we shall use \cite{janossy}, $^{Gd}H_{hf}$ = 10 $%
^{Y}H_{hf}$ and $\gamma _e$ = 1.6 $\times$ 10$^{4}$ $^{89} \gamma$. Using
the operating frequencies, $\nu$ = 15.64 MHz for $^{89}$Y NMR and $\nu$ =
245 GHz for Gd ESR, we can deduce the satellite separation from the main
line and the static and $T_1$ contributions to the linewidth at 100 K. These
are reported in Table 1. We find then that the expected width for the Gd ESR 
$nn$ line is at least four times larger than its shift, which justifies that
the satellite resonances are indeed quite difficult if not impossible to
observe.

The other issue is the suggestion by Janossy {\it et al.} that the
susceptibility corresponding to the pure YBCO$_7$ is restored at the Cu
neighbouring Zn. Since we observe a Curie-like increase of the $nn$ line
shift in Zn doped samples, increasing to values well above the YBCO$_7$
shift, the implication is that the hole content near the Zn dopants has not
been restored to that of undoped YBCO$_{7}$.

\subsubsection{NMR in Al doped La$_{2-x}$Sr$_x$CuO$_4$}

Recently, Ishida {\it et al.} \cite{ishida2} have performed NMR experiments
on La$_{2-x}$Sr$_{x}$CuO$_{4}$ (LASCO), in which non- magnetic Al is
substituted on the Cu site of the CuO$_{2}$ planes. Although they were
unable to detect the $nn$ nuclei of Al, they could directly detect the $%
^{27} $Al NMR signal. They did find that the shift of the $^{27}$Al NMR has
a Curie component. Since Al itself does not bear a local moment, this
observation can only result from a local moment which resides either on the $%
nn$ oxygen or copper orbitals which are coupled to the $^{27}$Al nuclear
spin via transferred hyperfine couplings. This observation does not enable
one to decide the location of the local moment. However, by analogy with our
results, the authors have inferred that it is located on the $nn$ Cu
orbitals. Their data are important as they confirm that a non-magnetic
substituent induces a local moment in a cuprate different from YBCO$_{6+x}$.
Ishida {\it et al.} \cite{ishida2} measured the $^{27}$Al NMR shift and $%
T_{1}$ which can be compared with the corresponding data on the $nn$ $^{89}$%
Y NMR in YBCO$_{6+x}$. As we stress below, several qualitative differences
in the experimental results are evident.

First, they analysed the $T$-dependence of the shift and susceptibility in
Al doped LASCO with a Curie-Weiss law with a sizeable Weiss temperature ($%
\theta $ $\approx $ 50 K). It is not so clear whether this high value of $%
\theta $ is also suggested by the $^{27}$Al NMR results since it might be
influenced by the reference taken for the $^{27}$Al chemical shift. In any
case, Ishida {\it et al.} do not demonstrate whether this large $\theta $
corresponds to a genuine single impurity effect or if it varies with Al
content thereby revealing a large coupling between the local moments. In
underdoped YBCO, we never found any indication for such a large deviation
from the Curie law, neither from NMR nor from susceptibility data. A
negative value $\theta $ $\approx $ -30K has been found by Monod {\it et al.}
\cite{monod} in YBCO:Zn by susceptibility measurements. However recent data
on samples with low content of impurity phases, by Mendels {\it et al.} \cite
{mendels1}, establish that a significant estimate of $\theta $ requires a
correct accounting of the susceptibility contribution of the pure compound.
They deduced $\theta $ $\simeq 4$ K for YBCO$_{6.64}$:Zn$_{4\%}$.

In order to analyse the $^{27}$Al relaxation rate data, Ishida {\it et al.}
use the simple local moment formulation of Eq. (7-10) (with different
notations). They provided a quantitative analysis of their data in which the
value of $J_{int}$ as deduced from their value of $\tau_{int}$ (a
microscopic probe) is fully consistent with their value of $\theta$ (from
bulk susceptibility, a macroscopic probe). This analysis would then appear
to be consistent with and support a simple local moment picture. We stress
here that not only do their results differ {\it qualitatively} from ours but
that an alternative interpretation is possible \cite{comment}. In their
analysis, they deduce a relaxation rate 1/$\tau$ which only varies slightly
with $T$. This would be expected if it were dominated by the $\tau_{int}$
term which is expected to be $T$-independent in the classical local moment
picture. Our results, on the other hand, have exhibited a totally opposite
trend with the $\tau_{ex}$ being negligible. We further find that in their $%
T_1$ analysis, Ishida {\it et al.}have taken a local moment Curie
susceptibility while their own $^{27}$Al data yielded a large Curie-Weiss
temperature ($\theta$ = 50 K). Introducing the actual 1/$(T + 50)$
dependence of $\chi$ in Eq. 7-10 lead us to deduce $1/(\tau_{ex}T)$ = 2 $%
\times$ 10$^{10}$ (sec K)$^{-1}$ which corresponds to a $T$ dependent
contribution to $1/\tau$ which is one order of magnitude larger than their
own result. This would yield a rather large value $J_{ex}$ = 0.17 eV which
contradicts their expectations.

However, such an analysis yields only a modest modification of the $T$%
-independent contribution which becomes $1/\tau _{int}$ = 5.6 $\times $ 10$%
^{12}$ sec$^{-1}$, only a factor of two smaller than their result which
still corresponds to a sizeable value of $\theta$, the Weiss temperature. Of
course, a significant difference between the two systems (YBCO$_{6.64}$ and
LASCO) is their doping range. While La$_{1.85}$Sr$_{0.15}$CuO$_{4}$ should
be considered close to an optimally doped material, in YBCO$_{6.64}$ we are
clearly in the underdoped regime which usually displays qualitatively
different properties. Unfortunately we do not have as complete results on
YBCO$_{7}$:Zn, which would allow for a direct comparison between the two
systems.

\subsection{The case of YBCO$_7$:Zn}

We have then clearly demonstrated that local moments are induced on Zn
substitution in YBCO$_{6.64}$. We have not studied samples with oxygen
contents other than 6.64 and 7 in great detail. However, we have seen, from
measurements on unoriented samples (see Fig.~12(b)), that the low-frequency
tail of the spectra which is associated with the outer satellite resonance
is less shifted from the main resonance position for increasing oxygen
content. We can then conclude that the local moment value decreases
gradually with increasing $x$. On reaching YBCO$_7$, we find that the $nn$
lines have practically merged with the main line. This is confirmed from
magnetisation data on impurity-phase free samples by Mendels {\it et al.} 
\cite{mendels1,mendels}, which show that the Curie constant for YBCO$_7$:Zn
is about one-sixth that of YBCO$_{6.64}$:Zn. Assuming the same hyperfine
couplings, the expected first $nn$ line position is shown in Fig.~ 21. In
view of the width due to a distribution of oxygen content, it is evident
that it will be difficult to resolve any extra resonance even for lower Zn
contents. Going to lower temperatures is ruled out as well due to the
relatively high $T_c$ of the samples.

We did not succeed either in distinguishing the $nn$ resonance from a
contrast of relaxation rate with the mainline. We shall see here that such a
contrast is not expected if we estimate the relaxation rate for the
outermost resonance by scaling the YBCO$_{6.64}$:Zn data at 100 K. The Curie
term in YBCO$_{7}$:Zn is about one-sixth that in YBCO$_{6.64}$:Zn and the
density of states at the Fermi level $\rho (\epsilon _{F})$ can be estimated
about 3 times higher, from the $^{89}$Y NMR shift data. Assuming that the
local moment to band coupling $J_{ex}$ stays unchanged, Equations 9 and 10
allow to deduce a contribution of the local moment fluctuations to 1/$T_{1}$
of $\approx $ 0.0018 sec$^{-1}$, which is much smaller than the observed
rate in undoped YBCO$_{7}$ (0.03 sec$^{-1}$). This confirms that for YBCO$%
_{7}$:Zn, local moment fluctuations are indeed difficult to detect on the $%
nn $ Y site.

In fact, the occurrence of a local moment in YBCO$_7$:Zn was evidenced first 
\cite{alloul2} from the presence of the oscillating long distance RKKY spin
polarisation of the host Cu spins. This was established from the Curie like
increase of the $^{89}$Y NMR linewidth observed in YBCO$_7$:Zn, and is
confirmed in the present experiments as well as from $^{17}$O NMR linewidth 
\cite{yoshinari} data.

\subsection{About AF correlations near the Zn impurities}

The experimental results on Gd ESR in YBCO and Al NMR in Al doped LASCO have
been interpreted along quite different lines. For Janossy {\it et al.}, the
absence of a Curie term for the ESR line shift led them to conclude that
there was no local moment. Instead, they considered that the susceptibility
of the YBCO system is restored near the Zn, as they found an increase of the
ESR shift with decreasing $T$. On purely experimental grounds it is not
clear whether the detected ESR signal involves all the Gd spins. We have
seen above that the outermost $nn$ resonances are not expected to be
resolved in the ESR data. However, the inner resonance might contribute to a
wing in the signal, and might explain the observed shift.

In any case, the present detailed $^{89}$Y NMR data demonstrate that the
central line is not shifted at all, so that the susceptibility is unmodified
at a few lattice distances from the Zn impurity. Second, the Curie-like
increase of the shift for the $nn$ reaches values well above those observed
for pure YBCO$_7$, which implies that the hole content near the Zn dopants
has not been restored to that of undoped YBCO$_{7}$. Finally, the bulk
susceptibility data (measured using a commercial SQUID magnetometer) of
Mendels {\it et al.} \cite{mendels1} indicate that the local moment
susceptibility increases down to 10 K, so that there is no doubt about the
occurrence of a Curie contribution. The fact that the $nn$ lines broaden
stongly with decreasing temperature is sufficient to explain that the Gd ESR
picks up only a part of the Gd signal which saturates at low- $T$. This is
somewhat reminiscent of the situation which prevailed in the preliminary $%
^{89}$Y NMR measurements done for large Zn concentrations in YBCO \cite
{alloul2}. In that case, the $nn$ resonances were not resolved and an
apparent shift of the $^{89}$Y NMR signal was observed. As the broadening of
the $nn$ lines is expected to be larger in the Gd ESR, the measured shift
involves a contribution of those $nn$ sites and is much smaller than that
expected for the 1$^{st}$ $nn$ signal.

As for the analysis of the Al NMR data in LASCO, the authors of course  do
not question the existence of a local moment. But, they still consider that
AF correlations are reduced near the impurity, and that the induced moments
on the four copper near neighbours are decoupled. Further, they even
anticipate that a state nearer to that observed in the overdoped material
prevails at distances just greater than the 1$^{st}$ $nn$ distance \cite
{ishida2}. However, we feel that there is no experimental evidence for such
a possibility in their work on the LASCO system. The only argument advanced
by Ishida {\it et al.} to support this hypothesis is the independent
experimental evidence found in their group for two $T_{1}$ components in
their $^{63}$Cu NQR measurements in Zn doped YBCO$_{7}$, and in YBa$_{2}$Cu$%
_{4}$O$_{8}$\cite{ishida1,zheng}. In both cases they find that the long $%
T_{1}$ component is longer than that observed in the pure system, and they
therefore associate it with Cu nuclei near the Zn impurities. This leads
them to conclude that the magnetic fluctuations near the Zn impurites have
been suppressed. However, the relative magnitude of the two components in
terms of number of sites and their dependence on Zn doping which could
support this interpretation has not been studied in detail. Most
importantly, the underlying idea seems to us to contain an essential
contradiction. Indeed if the AF fluctuations around the Zn were suppressed,
this would imply that Zn is in a classical metallic environment, which would
be totally inconsistent with the occurrence of a local moment \cite{comment}.

In the superconducting state, Ishida {\it et al.} do find $^{63}$Cu NQR
relaxation rates much larger than those found in the pure system, which
indicates the existence of states in the superconducting gap. This is also
seen from Yb Mossbauer experiments on samples in which a small fraction of Y
has been substituted by Yb \cite{hodges}. States in the gap in the
superconducting state induced by Zn impurities were also seen by neutron
scattering experiments \cite{sidis}. Those states are found at a scattering
vector of ($\pi$, $\pi$), even for YBCO$_7$:Zn , while a scattering at ($\pi$%
, $\pi$) for this oxygen content can hardly be detected in the pure system.
These experiments are thus direct evidences in favor of the persistence of
AF correlations in the vicinity of the impurities.

All these observations support the main point which we have been advocating,
that the AF correlations are at least maintained, and perhaps even
strengthened near the Zn impurities. In such a case the local moment cannot
be considered as formed of four independently fluctuating moments on the
four Cu sites $nn$ to Zn, but rather as an extended state involving further
neighbours, and in which the Cu $nn$ to Zn are ferromagnetically correlated
and fluctuate as a single entity.

We therefore think that the experimental observation done by Ishida {\it et
al.} on the normal state $^{63}$Cu NQR $T_1$ might have a quite distinct
interpretation. A more systematic study, possibly with different impurities
might be needed to clarify the origin of the longer $T_1$ component.

In conclusion, it seems to us that the existing experiments do not
contradict the main point of view originally proposed, i.e. that the local
moments induced by Zn are associated with the {\it correlated} nature of the
CuO$_2$ planes and that AF correlations might even actually be enhanced
around Zn.

\subsection{Induced spin polarisation at large distance from the Zn}

Up to now we have mainly considered the magnetic moments induced near the Zn
impurities. In noble metals hosts, any local charge perturbation is known to
induce long distance charge density oscillations (also called Friedel
oscillations). Similarly a local moment induces a long distance oscillatory
spin polarisation (RKKY) which has an amplitude which scales with the
coupling $J_{ex}$ of the local moment with the conduction electrons. This
oscillatory spin polarisation gives a contribution to the NMR shift of the
nuclei which decreases with increasing distance from the impurity. In very
dilute samples, if the experimental sensitivity is sufficient, the
resonances of the different shells of neighbours to the impurity can be
resolved \cite{alloul1}. These resonances merge together if the impurity
concentration is too large, which results then in a net broadening of the
host nuclear resonance.

Here, the occurrence of the local moment, even induced by the non-magnetic
substitution is also a local magnetic perturbation in the correlated host.
One therefore expects a response which will extend to long distances from
the impurity. Such contributions to the NMR linewidths have been found in
our work. We shall consider here in turn the case of YBCO$_{6.6}$ and that
of YBCO$_7$.

\subsubsection{YBCO$_{6.6}$}

Indeed, both the central $^{89}$Y line as well as the near neighbour
resonances have been found to be broadened in YBCO$_{6.6}$. As seen in
figures 7 and 9, these linewidths increase at low-$T$ and also increase with
increasing impurity content. The central line broadening is unfortunately
only a small fraction of the pure compound linewidth, and the temperature
dependence of the impurity induced contribution cannot be extracted
accurately. Experiments have therefore been performed by Bobroff {\it et al.}
\cite{bobroff} on $^{17}$O nuclei in substituted samples. The larger
hyperfine coupling of the $^{17}$O nuclei with the planar Cu as compared to
that of $^{89}$Y lead consequently to broadenings of the $^{17}$O NMR width,
which have been studied in great detail both for Ni and Zn substitutions. It
has been found that in both cases the broadening increases much faster than $%
1/T$ at low temperatures, contrary to what one might expect in a
non-correlated metallic host. This fast increase is a signature of the
anomalous magnetic response of the host which displays a peak near the AF
wavevector ($\pi$, $\pi$).

In the present experiments, the broadenings of the 1$^{st}$ $nn$ line
(Fig.~7) are somewhat related to this long distance polarisation induced by
the Zn impurity. The large increase of the $nn$ linewidth with increasing Zn
concentration is due to the distribution of susceptibility of the moments
associated with their mutual interaction. In a molecular field approach, the
Curie contribution $K_c$ to the shift of a $^{89}$Y $nn$ of a given Zn atom
is proportional to $\chi (H_0 + H_m)$, where $\chi$ is the single impurity
dimensionless susceptibility (= $c_{imp}/T$) and $H_m$ the molecular field
at the moment site induced by other Zn moments. This molecular field scales
with the magnetization of the local moments ($H_m$ = $kM$) and therefore
varies as $1/T$. The linewidth is then related to the root mean square value
of the molecular field $\delta H_m$. Consequently, the linewidth due to
interaction between the local moments (which scales with $\chi \delta H_m$)
should scale as $1/T^2$. We have therefore plotted in Fig.~ 22 the quantity $%
T^2 \Delta H_{corr}/H$ versus $T$, where $\Delta H_{corr} = \Delta H_{nn} -
\Delta H_{pure}$ is the increase of the $nn$ satellite linewidth (Full Width
at Half Maximum) with respect to that of $^{89}$Y in pure YBCO$_{6.6}$. We
can see that $T^2 \Delta H_{corr}/H$ is nearly $T$-independant as expected
from such a simple model. Let us point out that $H_m$ should in principle
behave as the long  distance spin polarisation detected by $^{17}$O NMR, and
should then increase faster  than $1/T$ at low temperature. Although the
experimental accuracy on the NMR width is  not great, a large increase of $%
T^2 \Delta H_{corr}/H$ is not observed at low $T$. More detailed and
possibly more accurate experiments are required to better understand whether
other contributions to the $nn$ linewidth have to be considered as well.

>From our data we can however get an overestimate of $\delta H_{m}$ from a
comparison of the magnitude of the linewidth with the actual shift of the $nn
$ line. Assuming a gaussian shape for this resonance $\Delta H_{corr}/2.36$
is simply proportional to $\chi \delta H_{m}$, while the shift $K_{c}H_{0}$
is proportional to $\chi H_{0}$. Therefore, $\delta H_{m}$ = 0.42 $\Delta
H_{corr}/K_{c}$. Further, from the analysis of Eq.~3, $K_{c}\simeq 0.024/T$.
From the discussion above, $\Delta H_{corr}$ = $2.36kc_{imp}^2H_0/T^2$ and
from Fig.~22 for 1 \% Zn, $\Delta H_{corr}\simeq 2H_{0}/T^{2}$. Then, for an
applied field $H_{0}$ = 7 Tesla, we deduce 
\begin{equation}
\delta H_{m}=250/T(Tesla/\%Zn)
\end{equation}
The molecular field becomes comparable with the thermal energy for $%
k_{B}T=\mu _{eff}H_{m}$, which for a measured $\mu _{eff}\simeq 0.8\mu _{B}$
in YBCO$_{6.6}$ corresponds to about 1.2 Tesla/K. Therefore the temperature
at which a spin-glass freezing of this system should occur can be estimated
to be about 15 K for 1 \% Zn. This number deduced from this rough analysis
is somewhat higher than that obtained from the Weiss temperature measured by
static susceptibility by Mendels {\it et al.} \cite{mendels1}, which does
not exceed 4 K for 4 \% Zn.  Apart from the above mentioned possible
experimental limitations, this difference could be linked with the fact that
we are dealing here with a 2D Heisenberg spin system, for which quantum
fluctuations reduce the spin-glass ordering temperature to $T_{g}=0$ \cite
{dekker}. A finite value for $T_{g}$ would then only result from weak
interplane  exchange couplings.

\subsubsection{YBCO$_7$}

The broadening has been found to increase as 1/$T$ at low-$T$, for this
slightly overdoped composition for which the planar susceptibility of the
pure system has little $T$-dependence. This increased linewidth at low-$T$
is a direct proof of the existence of a local moment behaviour induced by Zn
for this overdoped system \cite{alloul1}. We have seen that the absence of
near neighbour resonance lines is also an indication that the effective
moment is very small, which confirms the susceptibility data of Mendels {\it %
et al.}. A similar observation has been made by Bobroff {\it et al.} \cite
{bobroff} through $^{17}$O NMR data. Initially such a broadening can be
explained with the RKKY-like broadening induced by the local moments.
However, in the underdoped case it has been shown that the response of the
correlated electronic system is quite distinct from that of a free electron
gas, as the $^{17}$O NMR linewidth  exhibits an anomalous $T$ dependence The
NMR data for the YBCO$_7$ composition, both for $^{17}$O and $^{89}$Y, does
not display such an anomalous T dependence, and one might wonder whether a
conventional RKKY broadening is then recovered. Such an approach has been
used in our initial report \cite{mahajan}. However, the very fact that a $T$%
- dependent magnetic behaviour is induced by Zn substitution is an
indication that {\it the correlated nature of the electronic state has not
disappeared in} YBCO$_7$. This is also established by the well known
anomalous non-Korringa $T$-dependence of $^{63}$Cu \cite{takigawa}.
Therefore a direct test of the shape of the spatial dependence of the
impurity induced spin polarisation should give us information on the
importance of these correlations. This aim will be pursued in the future
with careful studies of the NMR lineshapes, which are expected to be more
sensitive to the detailed shape of the spin polarisation \cite{bobroff}.
Such experimental studies will be undertaken on $^{17}$O NMR which possesses
a larger signal to noise ratio. A comparative discussion of the induced spin
polarisation as sensed by the $^{89}$Y and the $^{17}$O nuclei will
therefore be performed in the future.

\section{Conclusions}

A large variety of conclusions have been drawn and various questions have
been raised from the present results. They address different points
extending from the materials properties to detailed questions on the
electronic structure of the impurities and their influence on
superconductivity.

First, concerning the {\it physical chemistry} of the cuprates, the
intensity of the near neighbour resonances allowed us to calibrate the
amount of Zn substituted on the planar Cu sites. Our result is the strongest
experimental proofthat the Zn substitutes dominantly on this planar site, up
to 3 \% Zn, and within 10 \% experimental accuracy.

We have confirmed the influence of Zn impurities on the {\it phase diagram }
of the cuprates in the underdoped regime. The implication that the static
and dynamic susceptibility far from the impurity is unaffected by Zn is
borne out by our shift and relaxation data. This demonstrates that the
related {\bf q} = 0 pseudo-gap is not modified. The change of the
macroscopic susceptibility is only associated with modifications of magnetic
properties in the vicinity of the impurity. Kakurai {\it et al.} \cite
{kakurai} initially suggested, on the basis of their neutron scattering
experiments, that the pseudo-gap vanishes at {\bf q} =$(\pi ,\pi )$ while
the gap at other {\bf q} values is unchanged. However, the neutron data of
Sidis {\it et al.} \cite{sidis} in fact suggests that the pseudo-gap at {\bf %
q} =$(\pi ,\pi )$ does not vanish but that some states appear in the
pseudo-gap. Those could also be associated with the local magnetic
modifications induced around the Zn. In a scenario in which the pseudo gaps
would be associated with the formation of local pairs at high-$T$, these
results ndicate that impurities do not prevent the formation of local pairs
except possibly in their vicinity.

What are the actual magnetic properties {\it in the vicinity of the Zn
impurity}? Although our early experiments had given strong proofs of the
occurrence of a local moment behaviour induced by non-magnetic Zn
impurities, the validity of this observation has been periodically put into
question. The significance of the $nn$ $^{89}$Y NMR results has been, for
instance, questioned from the absence of detectable $nn$ resonances of Zn in
the ESR experiments on Gd/Y substituted underdoped samples. We have clearly
shown here that the large expected relaxation rate induces a broadening of
the Gd ESR $nn$ lines which prohibits their detection. The authors have also
concluded from those Gd ESR experiments that the full density of states
corresponding to pure YBCO$_{7}$ is restored near the Zn impurity. The fact
that the $^{89}$Y 1$^{st}$ $nn$ resonance is found to display an NMR shift
much larger than that of the optimally doped compound at low- $T$, is clear
evidence against this idea. On the contrary, the susceptibility of the Cu $%
nn $ to Zn is found to present a Curie like $T$-dependence, hence the
``local moment'' denomination, which we have been using throughout. This
local moment behaviour is confirmed in YBCO$_{6.6}$ by macroscopic
susceptibility SQUID data \cite{mendels1,mendels}.

It is clear that the observed local moment behaviour is original inasmuch as
it is the {\it magnetic response of the correlated electron system to the
presence of a spinless site.} The perturbation induced by Zn extends at
least to the four $nn$ copper sites, but we have shown that, in underdoped
YBCO$_{6.6}$, our data are compatible with a local dynamic AF state which
extends over more Cu sites. Although the present NMR data are not sufficient
to allow us to determine the actual extension of this state, the width of
the neutron scattering peak at ($\pi$, $\pi$) which is found to develop at
low-$T$ within the pseudo-gap in presence of Zn \cite{sidis}, corresponds to
a real space extension of at least 7 \AA.

Various theoretical arguments in favor of the {\it occurence of a local
moment in presence of a spinless site in a correlated electronic system}
have been advanced \cite{fink,nagaosa,poilblanc,khaliullin}. As complete
understanding of the magnetic properties of pure cuprates is far from being
achieved, it is no surprise that the present theoretical descriptions of the
impurity induced magnetism are rather crude, and for example, do not address
its microscopic extent. Our results might, however, be put in parallel with
recent theoretical work on undoped quantum spin systems. For instance
Martins \cite{dagotto} predicts static local moments induced due to doping $S
$ = 1/2 Heisenberg AF chains or ladders with non-magnetic impurities. NMR
experiments on the S = 1/2 Heisenberg chain system Sr$_{2}$CuO$_{3}$ are
consistent with the prediction of an induced local moment with a large
spatial extent along  the chain \cite{takigawa2}. In this undoped insulating
quantum liquid, the response is then purely magnetic. Since the parent
compound to YBCO superconductors is a 2D Heisenberg AF and dynamic AF
correlations appear to persist even in the metallic compositions, appearance
of local moments on many Cu sites near to the doped Zn might well be
anticipated.

In the slightly overdoped YBCO$_{7}$, the local moment could initially only
be detected through the induced long distance spin polarisation \cite
{alloul2}. A local moment induced by non-magnetic Al substituted on Cu is
also detected in optimally doped LASCO from $^{27}$Al NMR . The fact that we
could not resolve the $nn$ signal in YBCO$_{7}$ is consistent with the weak
magnitude found for the Curie like contribution to the local susceptibility.
The {\it decreasing magnitude of the moment with increasing hole doping}
could carefully be monitored by direct SQUID measurements \cite{mendels}.
This decrease could be linked experimentally with a {\it decreasing
screening radius}by the conduction band. However, the magnetic states which
are detected within the spin-gap at low-$T$ by neutron scattering \cite
{sidis} exhibit a short magnetic correlation length, so that the spatial
extent of the local moment also decreases with increasing hole doping as
does the AF correlation length in the pure system. Altogether, our
experiments cannot, at present, distinguish the {\it respective roles of the
screening radius and the AF correlation length in defining the local moment
magnitude and spatial extent}.

Another important question which arises then concerns the {\it coupling of
the defect local moment to the host}. For magnetic impurities in simple
metals, an exchange coupling $J_{ex}$ between the local moment and the
conduction electron spins usually occurs, and determines some of the
thermodynamic properties of the local moment. For instance the {\it %
fluctuation rate of the local moment} ($1/\tau $) is directly determined by $%
J_{ex}$, and follows a Korringa relation in classical metals. This
fluctuation time can be estimated from nuclear spin lattice relaxation data.
In the YBCO system we could only obtain such measurements in the underdoped
regime on the $^{89}$Y $nn$ of Zn. From these we could show that only weak
contributions to $1/T_{1}$ are expected on the $^{89}$Y $nn$ in the
optimally doped case, and could not be sensed within experimental accuracy.
Direct measurements of the $^{27}$Al $T_{1}$ are on the contrary sensitive
enough in the optimally doped case in LASCO, as seen by Ishida {\it et al}.
Their results, although they establish the occurrence of a local moment
induced by the spin-less Al$^{3+}$ substituent, differ markedly from those
obtained in YBCO. A large temperature independent contribution to the shift
and local moment fluctuation time is detected, contrary to our observations.
The origin of the difference might be

i) linked with the larger valence of Al$^{3+}$, i.e. the charge difference
with respect to host planar Cu$^{2+}$,

ii) a peculiarity of the LASCO system, as indeed the physical properties of
this system do not appear to fit in a universal picture with the other
cuprates (see Ref.~\cite{bobroff1}),

iii) or merely a difference between underdoped and optimally doped systems,
as the experiments could not be performed on the two systems under similar
conditions.

Further experiments will certainly permit to make a decision between these
possibilities. Currently, experiments do not permit a clear indication on
the applicability of an exchange model. In conventional metallic systems,
the local moment couples through $J_{ex}$ to the electron bath and an
oscillatory RKKY polarization occurs in the band. Therefore $J_{ex}$ can be
usually estimated from the broadening of the host NMR \cite
{alloul3,walkerwalstedt}.Applying the standard RKKY theory yields values of
the exchange coupling which are very large \cite{mahajan}. But, we have
recently shown in Orsay \cite{bobroff} that, at least in the underdoped
regime, the behaviour of the $^{17}$O linewidth does not follow the expected
RKKY $T$ dependence at all, i.e. the NMR width does not scale solely with
the impurity magnetization. Let us note here that whatever the method used 
\cite{mahajan}, \cite{ishida2}, the estimates of the coupling constant are
presently such that if one applies a simple exchange model, one would expect
a {\it large Kondo temperature} $T_{K}$ and correspondingly, a spin
susceptibility which would deviate from the Curie dependence at $T$ $\sim $ $%
T_{K}$ and saturate below. From SQUID data, Mendels {\it et al.} concluded
that this is not the case, and that $T_{K}$ does not exceed at most a few K,
in the underdoped YBCO compounds. Such a Kondo-like effect was a candidate
mechanism for the reduction of the magnitude of the local moment in YBCO$_{7}
$:Zn (see for instance Ref. \cite{nagaosalee}). But obviously the Kondo
model needs to be revised in the context of a strongly correlated electron
system. Such difficulties had been already pointed out by Hirschfeld \cite
{hirschfeld} in view of our preliminary experimental results.

In conclusion, we have detailed here the experimental evidence for the
occurence of a local moment behaviour induced by spinless substitutions on
the Cu site in CuO$_{2}$\ planes of cuprates. The existence of original
magnetic behaviour induced by non-magnetic substitutions can be anticipated
from current theoretical treatments of {\it undoped} low-dimensional spin
systems. However, the detailed experimental observations reported here on 
{\it doped} cuprates do not have a thorough interpretation from the
theoretical standpoint. We suggest that further experimental and theoretical
efforts regarding these properties are essential to lead us towards a
comprehensive description of the magnetic and superconducting properties of
the cuprates.

We would like to thank P. Mendels, J. Bobroff, and A. Macfarlane for useful
discussions and comments about the manuscript.

Laboratoire de Physique des Solides is a ``Unit\'{e} Mixte de Recherches du
Centre National de la Recherche Scientifique et de l'Universit\'{e}
Paris-Sud''.

\newpage

{\bf Figure Captions}

FIG.~1 $^{89}$Y NMR lineshape at 130 K in YBCO$_{6.64}$:Zn$_{1\%}$ when the
duty cycle of the pulse sequence, $t_{rep}$, is 20 sec, with the sample $c$
axis aligned parallel or perpendicular to the applied field. The relative
intensity of the $nn$ lines is enhanced here, since they have a $T_1$
comparable to $t_{rep}$ while the mainline $T_1$ is much longer.

FIG.~2 $^{89}$Y NMR lineshapes at 100 K in YBCO$_{6.64}$:Zn$_{y\%}$, for the
sample $c$ axis aligned parallel to the applied field $H$. The relative
intensities of the outer and middle lines are seen to qualitatively increase
with increasing $y$.

FIG.~3 (a) $^{89}$Y NMR lineshape at 90 K in YBCO$_{7}$:Zn$_{1\%}$ showing
absence of resolved $nn$ lines indicating that the induced local moment
magnitude is weak in YBCO$_7$:Zn as compared to that in YBCO$_{6.64}$:Zn.
Also shown is the decomposition of the lineshape for YBCO$_{6.64}$:Zn$_{1\%}$
into three gaussians.

FIG.~3 (b) $^{89}$Y NMR lineshape at 80 K in YBCO$_{7}$:Zn$_{2\%}$ for
different repetition rates. The arrow indicates the position of the middle
line in YBCO$_{6.64}$:Zn. The lineshape stays nearly unchanged indicating
the absence of any components relaxing faster than the mainline.

FIG.~4 Fully relaxed $^{89}$Y spectra in YBCO$_{6.64}$:Zn$_{y\%}$. The solid
lines are fits to three gaussians as explained in the text. (a)1\% Zn (b)2\%
Zn (c)4\% Zn.

FIG.~5 Variation of the fractional $nn$ line intensity (integrated) as a
function of Zn content $y$ for YBCO$_{6.64}$:Zn$_{y\%}$. The solid lines
correspond to variations as expected from statistical models as explained in
text. The intensity of the outermost line is seen to be in near agreement
with that expected from a $^{89}$Y nuclei nearest to the doped Zn. The
middle line intensity might agree with that expected from a combination of 2$%
^{nd}$ and 3$^{rd}$ $nn$ $^{89}$Y nuclei.

FIG.~6 $^{89}$Y $nn$ line shifts $K$ versus temperature $T$ for YBCO$_{6.64}$%
:Zn$_{y\%}$. The outer line shift is seen to have a strong upturn with
decreasing $T$ indicating a coupling to a Curie-like susceptibility. The
solid lines are drawn as guides to the eye.

FIG.~7 Linewidths normalized to the applied magnetic field $\Delta H/H$ of
the $nn$ lines for YBCO$_{6.64}$:Zn$_{y\%}$ are seen to increase with
decreasing $T$. The solid lines are drawn as guides to the eye.

FIG.~8 $^{89}$Y mainline shift $K$ versus temperature $T$ for YBCO$_{6.64}$%
:Zn$_{y\%}$ is nearly unchanged from that of YBCO$_{6.64}$ indicating little
change in the hole content with Zn doping.

FIG.~9 Normalized linewidth $\Delta H/H$ of the mainline versus the
temperature $T$ for YBCO$_{6.64}$:Zn$_{y\%}$. Unlike YBCO$_7$ which has a
nearly $T$ independent linewidth, YBCO$_{6.64}$ linewidth decreases below
120 K and increases again at much lower temperatures. Qualitatively, the
linewidths for Zn doped YBCO$_{6.64}$ are higher than the undoped compound.
The solid lines are drawn as guides to the eye.

FIG.~10 The $T$ variation of $^{89}$Y mainline shift $K$ for YBCO$_{7}$:Zn$%
_{y\%}$ is unchanged from that in YBCO$_7$ again indicating that the hole
content is nearly unchanged with Zn doping. Note that the chemical shift
reference is about 150 ppm, hence the change in susceptibility on Zn
addition is at most 4 \% of the susceptibility.

FIG.~11 Normalized linewidth $\Delta H/H$ of the mainline versus the
temperature $T$ for YBCO$_{7}$:Zn$_{y\%}$. The linewidth increases in a
Curie-like manner with decreasing $T$. Also, the linewidth is larger for
larger Zn contents. Solid lines are drawn as guides to the eye.

FIG.~12(a) $^{89}$Y mainline shift $K$ versus temperature $T$ for YBCO$%
_{6+x} $:Zn$_{4\%}$. The $T$-dependence is similar to that of YBCO$_{6+x}$.

FIG.~12(b) Variation of lineshape as a function of $x$ for YBCO$_{6+x}$:Zn$%
_{4\%}$. The low frequency tails which correspond to the shifted satellite
lines are seen to appear already for oxygen content $x$ = 0.84.

FIG.~13 Spectra for YBCO$_{6.64}$:Zn$_{1\%}$ for various $t_{rep}$ values.
It is clear that the outer line recovers its full intensity for much smaller 
$t_{rep}$ values than the mainline and hence has a much shorter $T_1$ than
the mainline.

FIG.~14 Analysis of the relaxation rate data of Fig.~13 for YBCO$_{6.64}$:Zn$%
_{1\%}$. The data have been fitted to a sum of three gaussians and the
deduced intensities corresponding to the main and the $nn$ lines are plotted
versus the repetition time of the pulse sequence $t_{rep}$ on a semi-log
scale. The solid lines are fits to a single exponential recovery.

FIG.~15 (a) $^{89}$Y nuclear spin-lattice relaxation rate divided by
temperature 1/$T_1T$ versus temperature $T$ for the mainline and the middle
satellite in YBCO$_{6.64}$:Zn$_{y\%}$. (b) these data are for the outermost
satellite line. The $nn$ lines are seen to have a shorter $T_1$ than the
mainline.

FIG.~16 Nuclear magnetization corresponding to the mainline versus the
repetition time of the pulse sequence $t_{rep}$ for YBCO$_{7}$:Zn$_{y\%}$.
The fact that the data can be fit to a single exponential (solid line)
indicates the absence of any other components to the relaxation.

FIG.~17 $^{89}$Y nuclear spin-lattice relaxation rate divided by temperature
1/$T_1T$ versus temperature $T$ for YBCO$_{7}$:Zn$_{y\%}$. Note the
magnified scale of the $y$ axis. As for YBCO$_{6.64}$:Zn, the $T_1$ of the
mainline is not affected.

FIG.~18(a) Curie component (in addition to a constant) of the $^{89}$Y shift 
$K$ versus temperature $T$ for the outermost line in YBCO$_{6.64}$:Zn$_{1\%}$%
. The solid line is a fit (see Eq. (2)) assuming that the local moment is on
the oxygen atoms. (b) The solid line is a fit (see Eq. (3)) assuming moments
on the copper atoms.

FIG.~19 A schematic of the location and the orientation of the local
magnetization around a Zn impurity.

FIG.~20 The ratio of the relaxation rate of the local moment spin $1/\tau $
to temperature $T$ for the 1$^{st}$ $nn$ $^{89}$Y nuclei in YBCO$_{6.64}$
:Zn. $\tau $ has been determined from Eq. (10).

FIG.~21 $^{89}$Y NMR spectrum for YBCO$_{7}$:Zn$_{y\%}$. Expected position
of outermost line, based on the macroscopic susceptibility data (see text)
is indicated by an arrow. The actual linewidth even in the pure compound is
such that any feature at this position cannot be resolved.

FIG.~22 Variation of the product of the square of temperature $T^2$ and the $%
nn$ linewidth $\Delta H_{corr}$ (linewidth of the pure compound has been
subtracted from the measured linewidth) with the temperature $T$ for YBCO$%
_{6.64}$:Zn.

\begin{table}[h]
\caption{$^{89}$Y NMR measured values (at about 100 K) of the separation of
the 1$^{st}$ $nn$ line from the mainline $\protect\delta \protect\nu$, the
spin- lattice relaxation rate of the 1$^{st}$ $nn$ $1/T_1$, and the
linewidth of the 1$^{st}$ $nn$ $\Delta \protect\nu$ are listed along with
calculated values (see text) of the same for Gd ESR in YBCO$_{6.64}$:Zn.}
\begin{tabular}{ccc}
Parameter & $^{89}$Y NMR & Gd$^{3+}$ ESR \\ \hline
$\delta \nu$ (Hz) & 4 $\times$ 10$^3$ & 6.4 $\times$ 10$^8$ \\ 
$1/T_1$ (Hz) & 0.1 & 2.56 $\times$ 10$^9$ \\ 
$\Delta \nu$ (Hz) & 2 $\times$ 10$^3$ & 2.56 $\times$ 10$^9$%
\end{tabular}
\end{table}

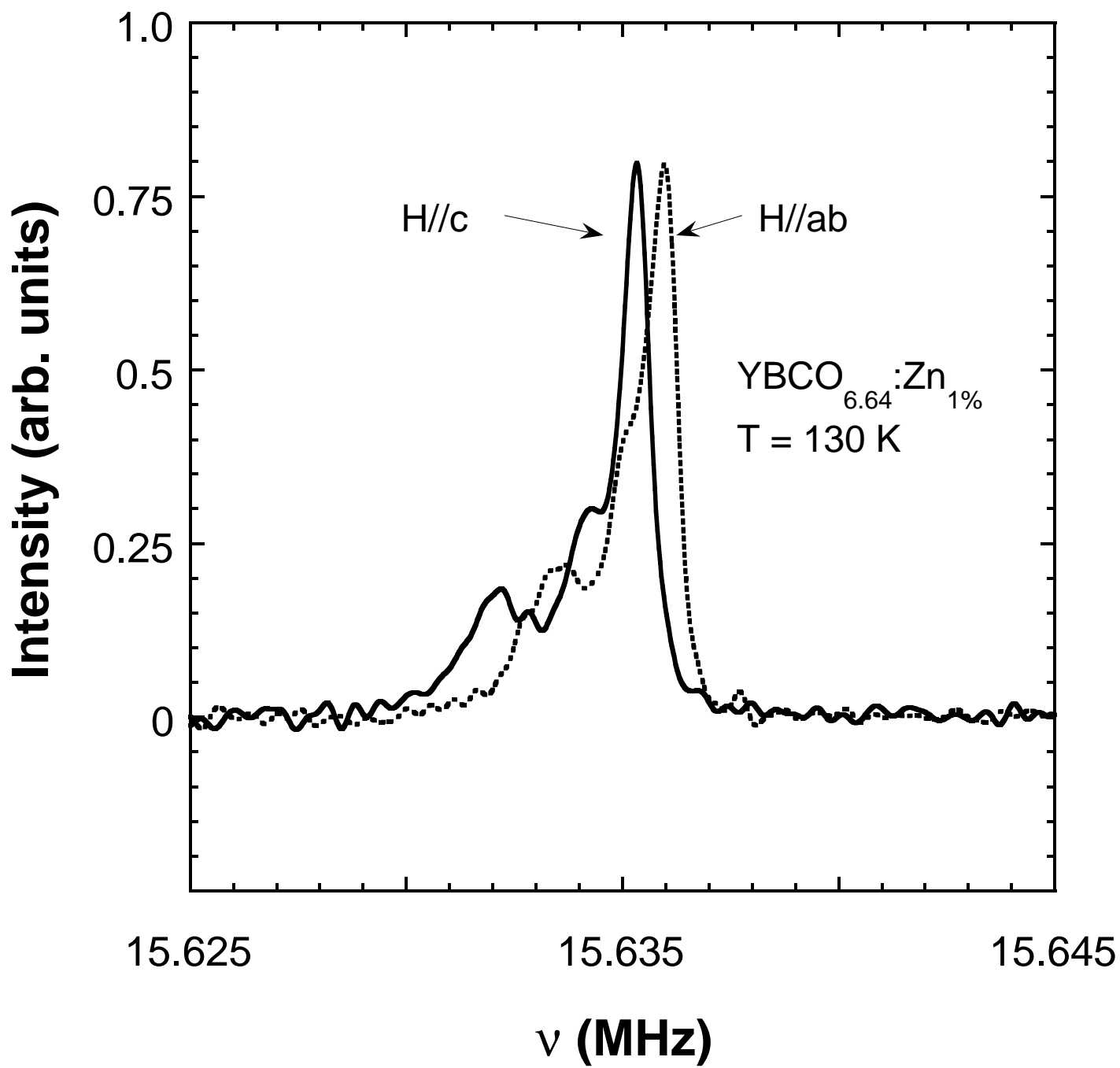
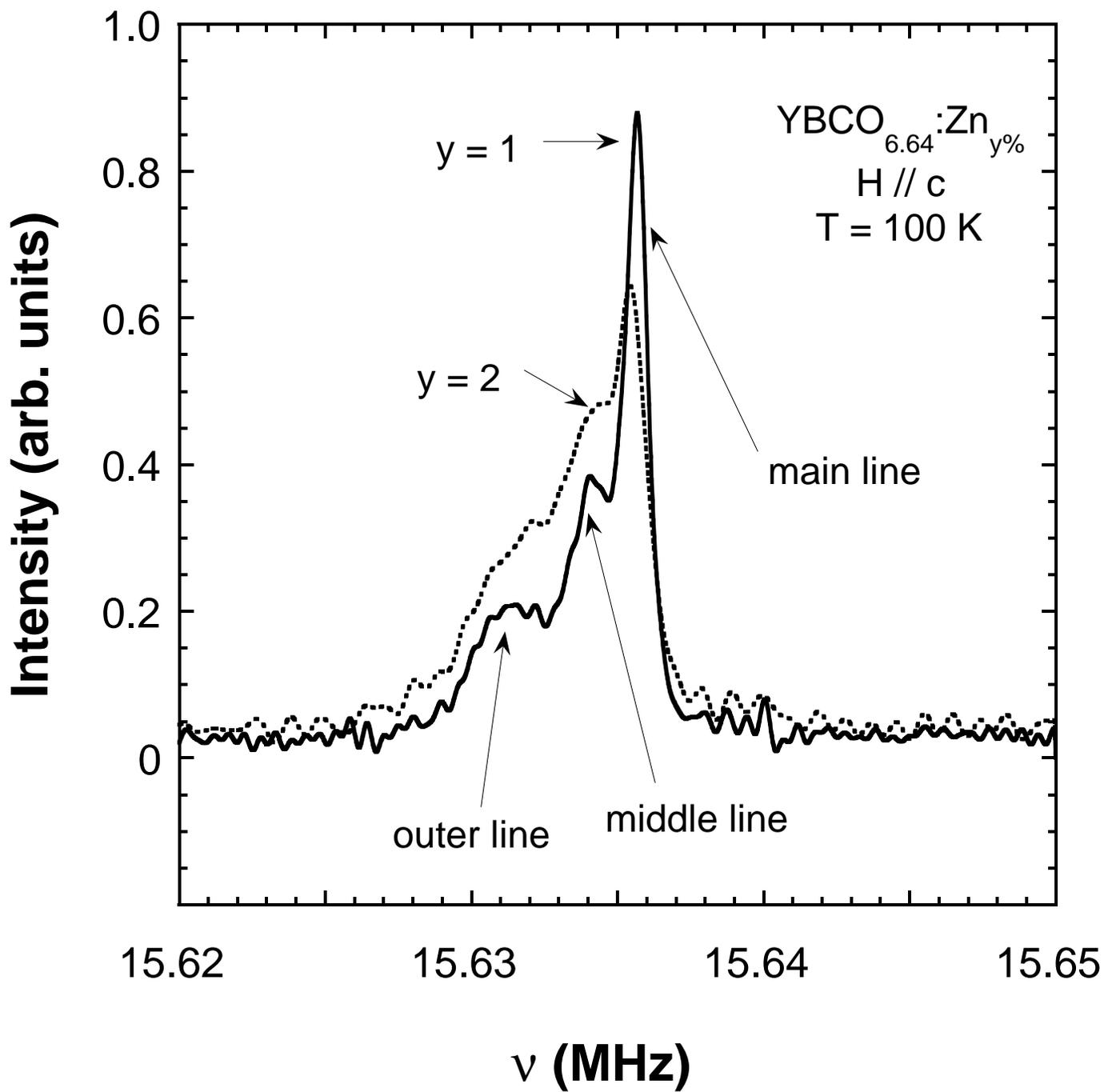
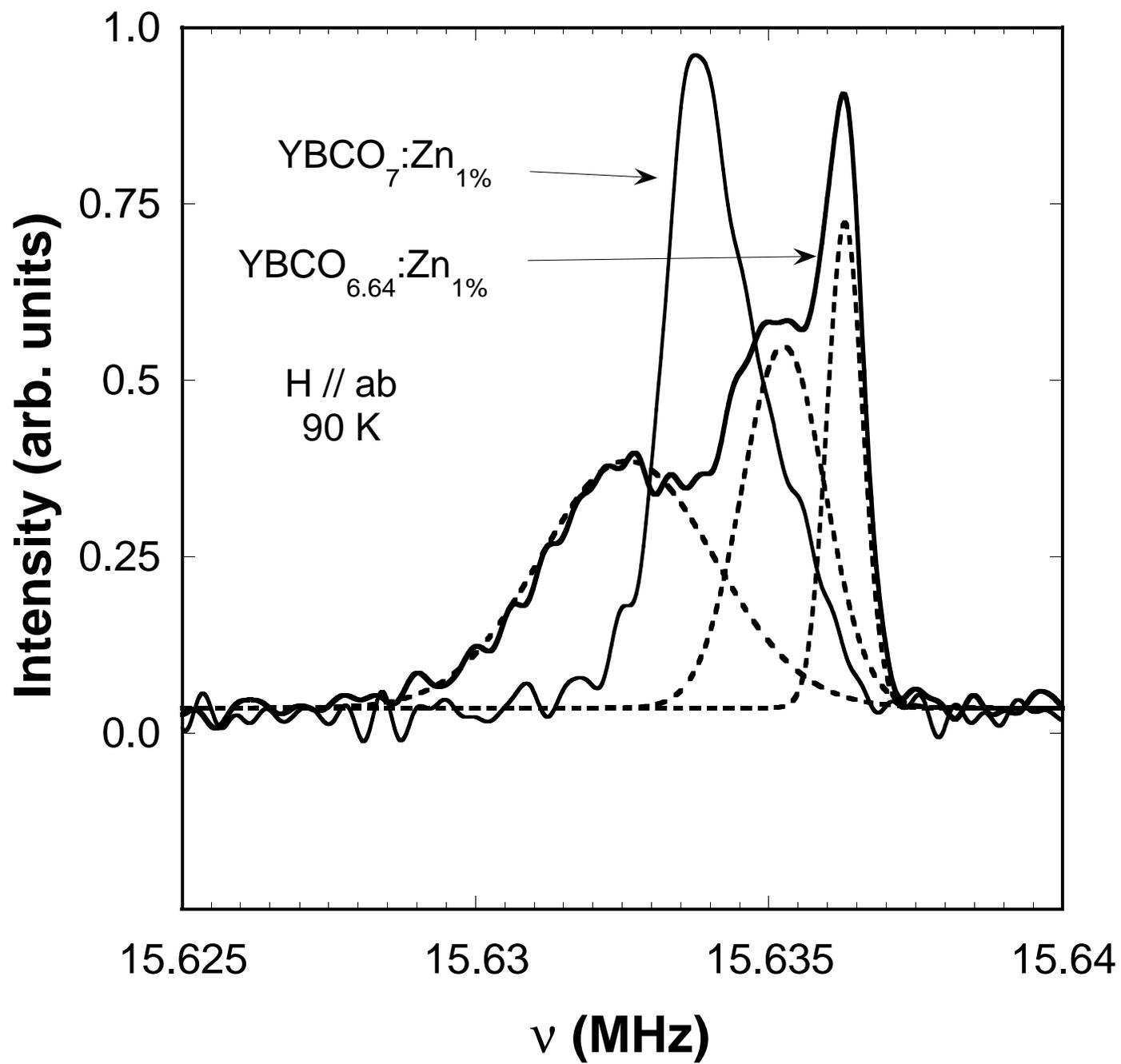
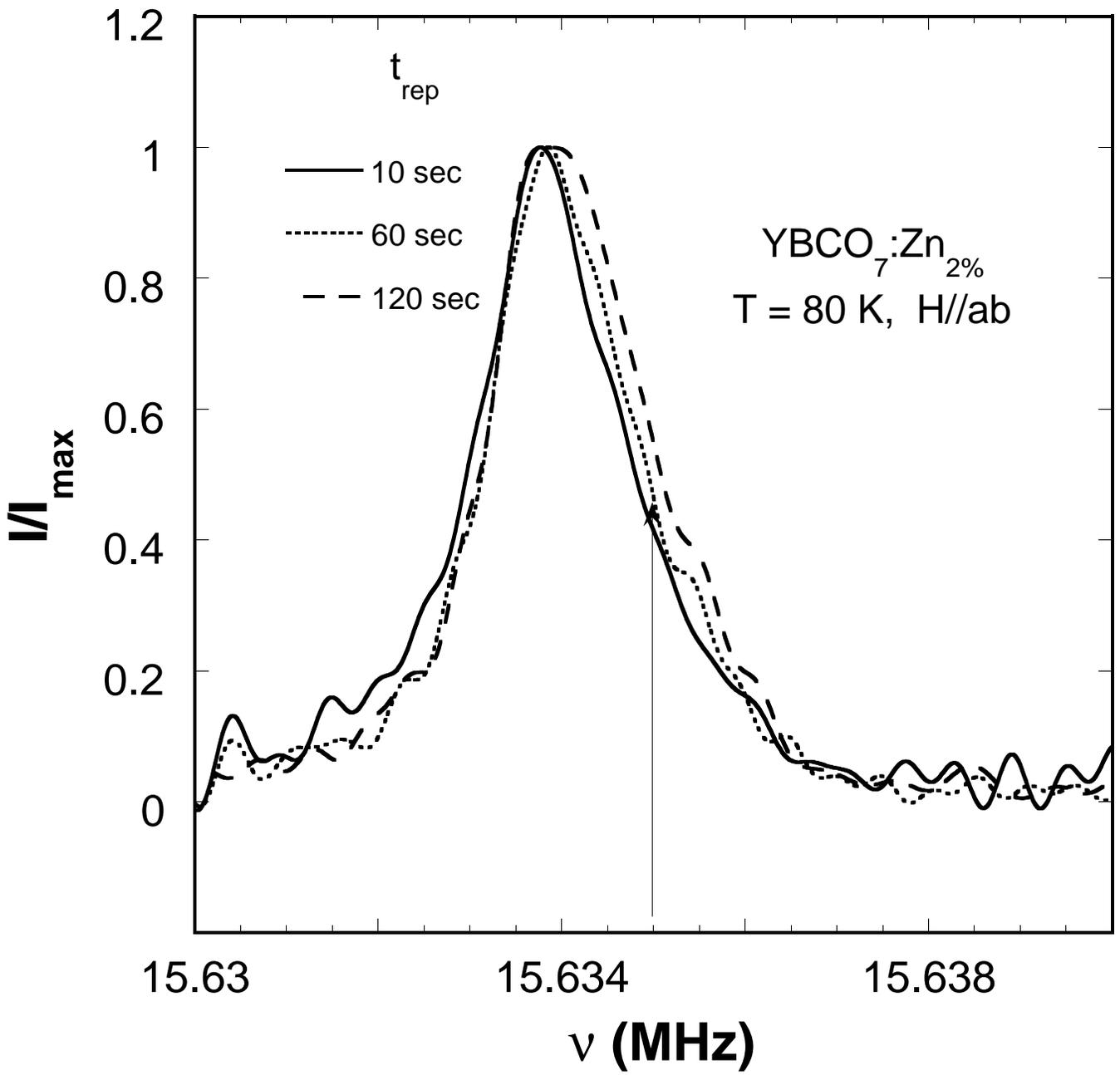
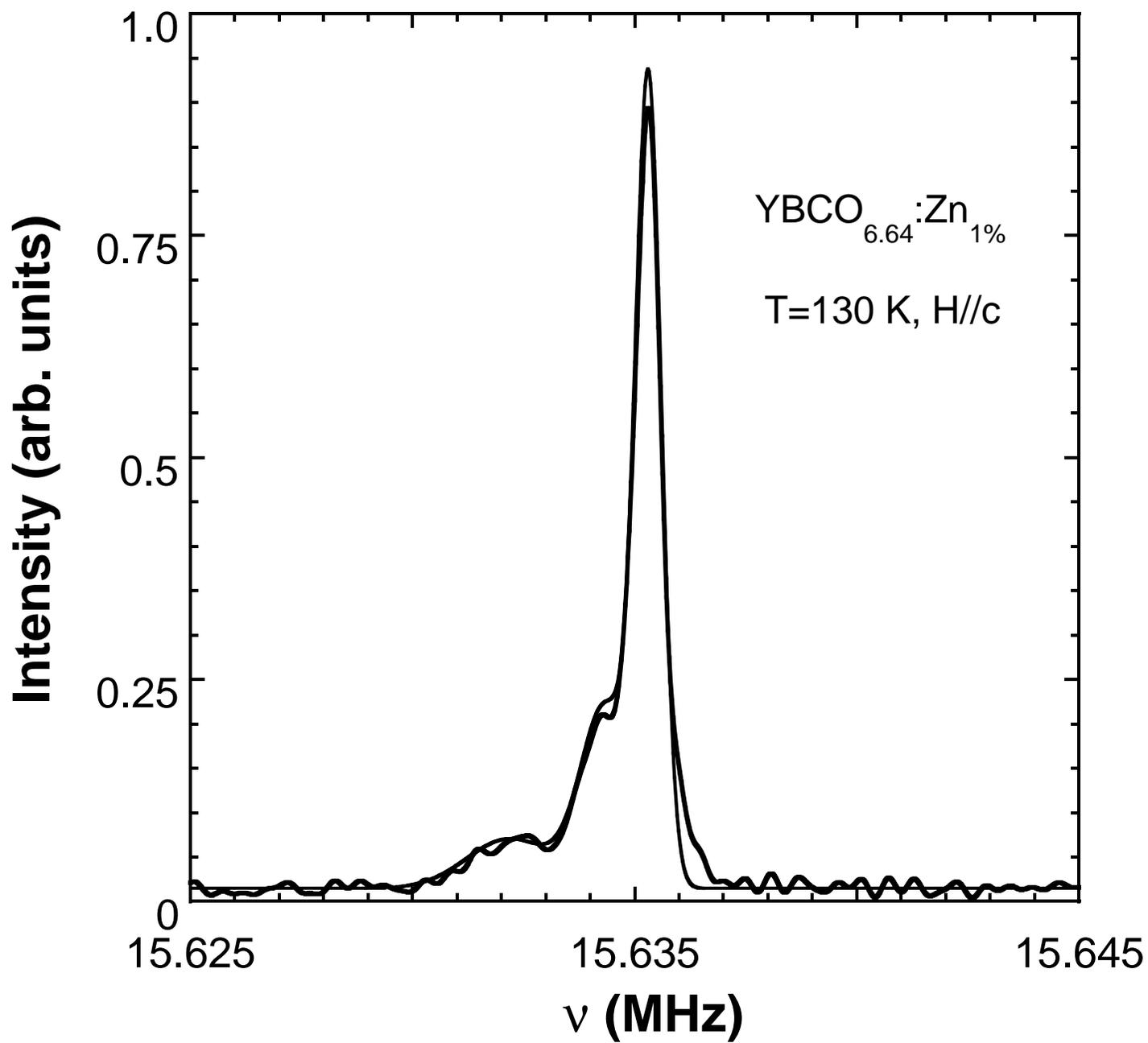
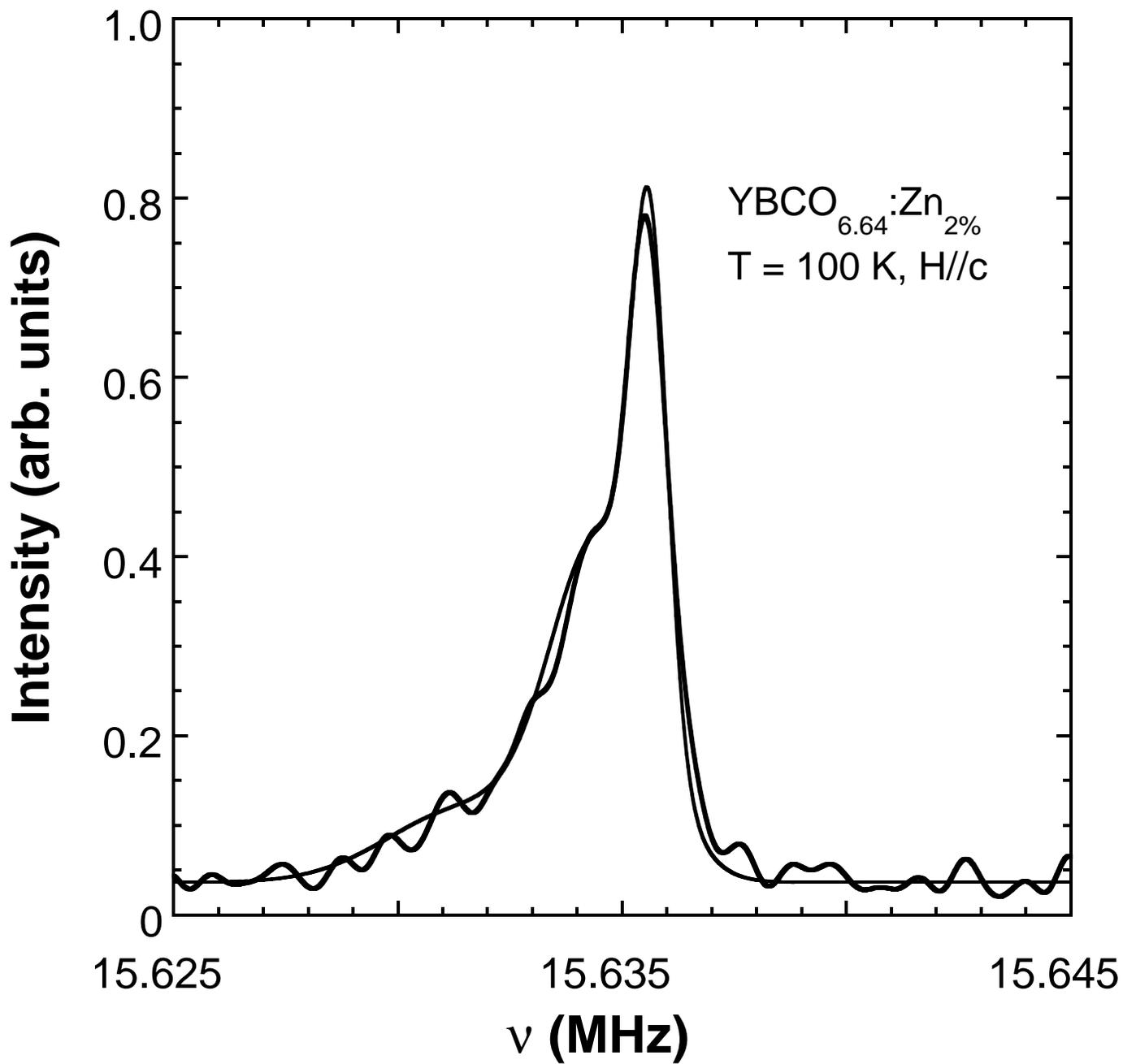
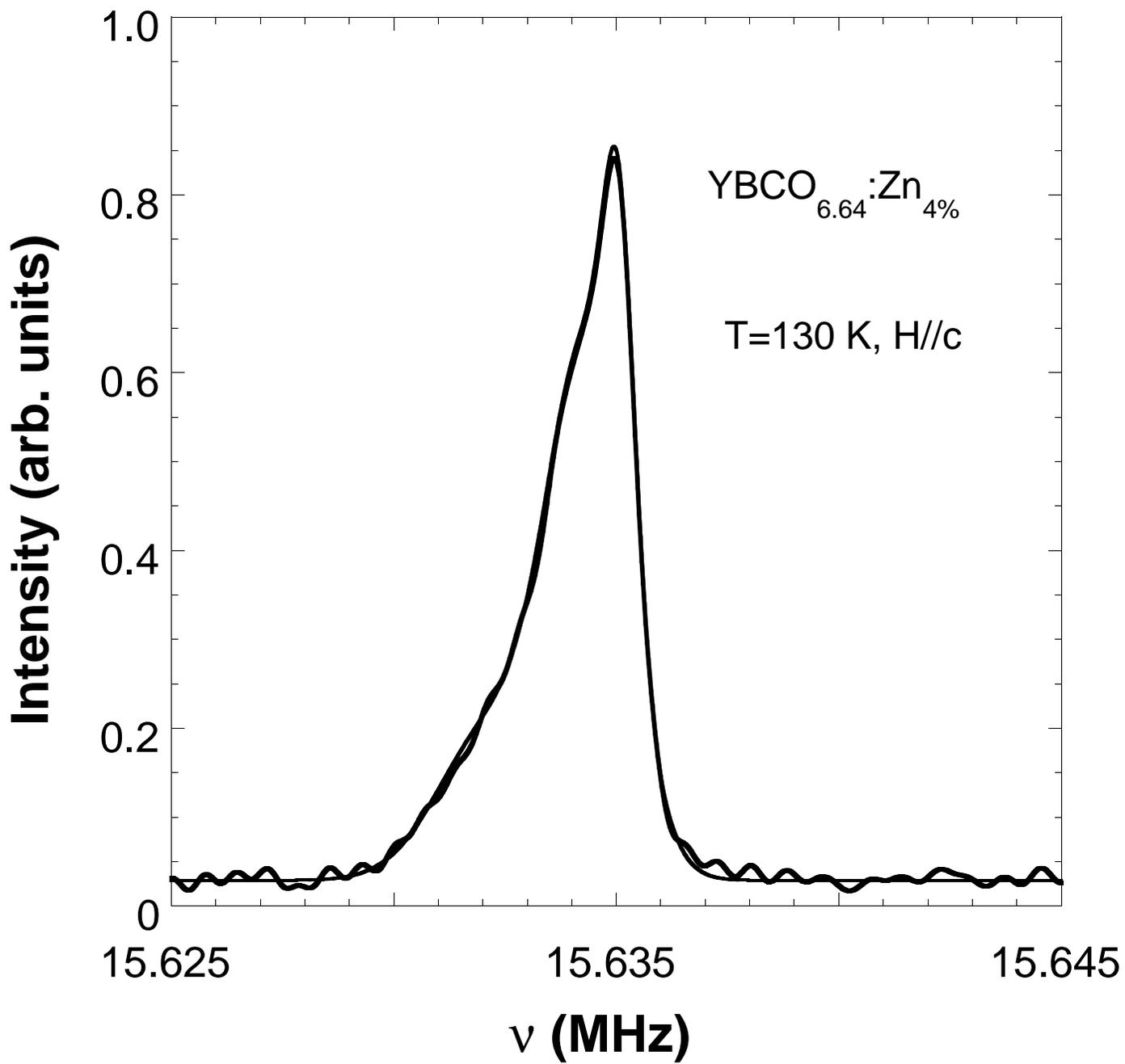
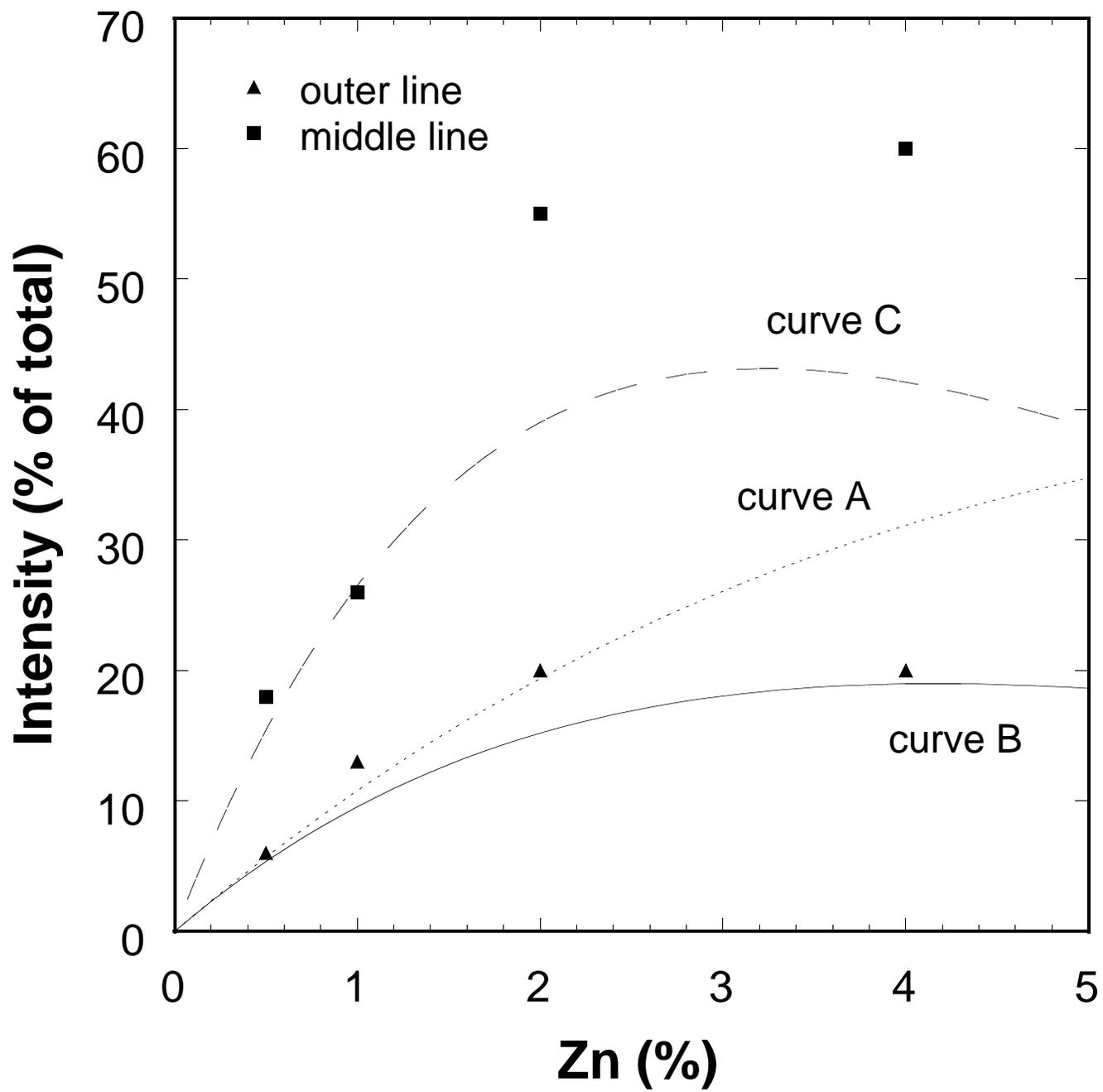
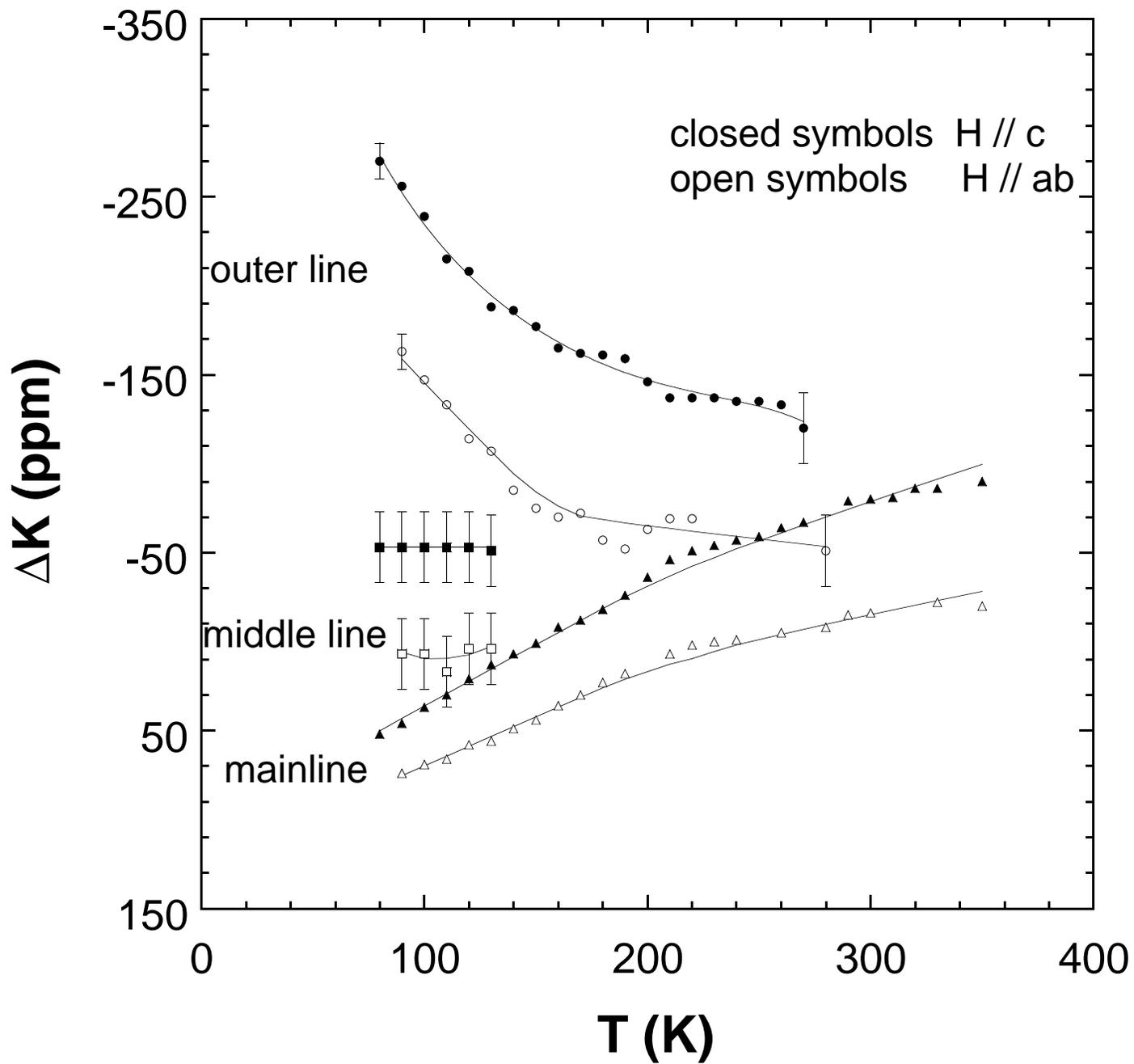
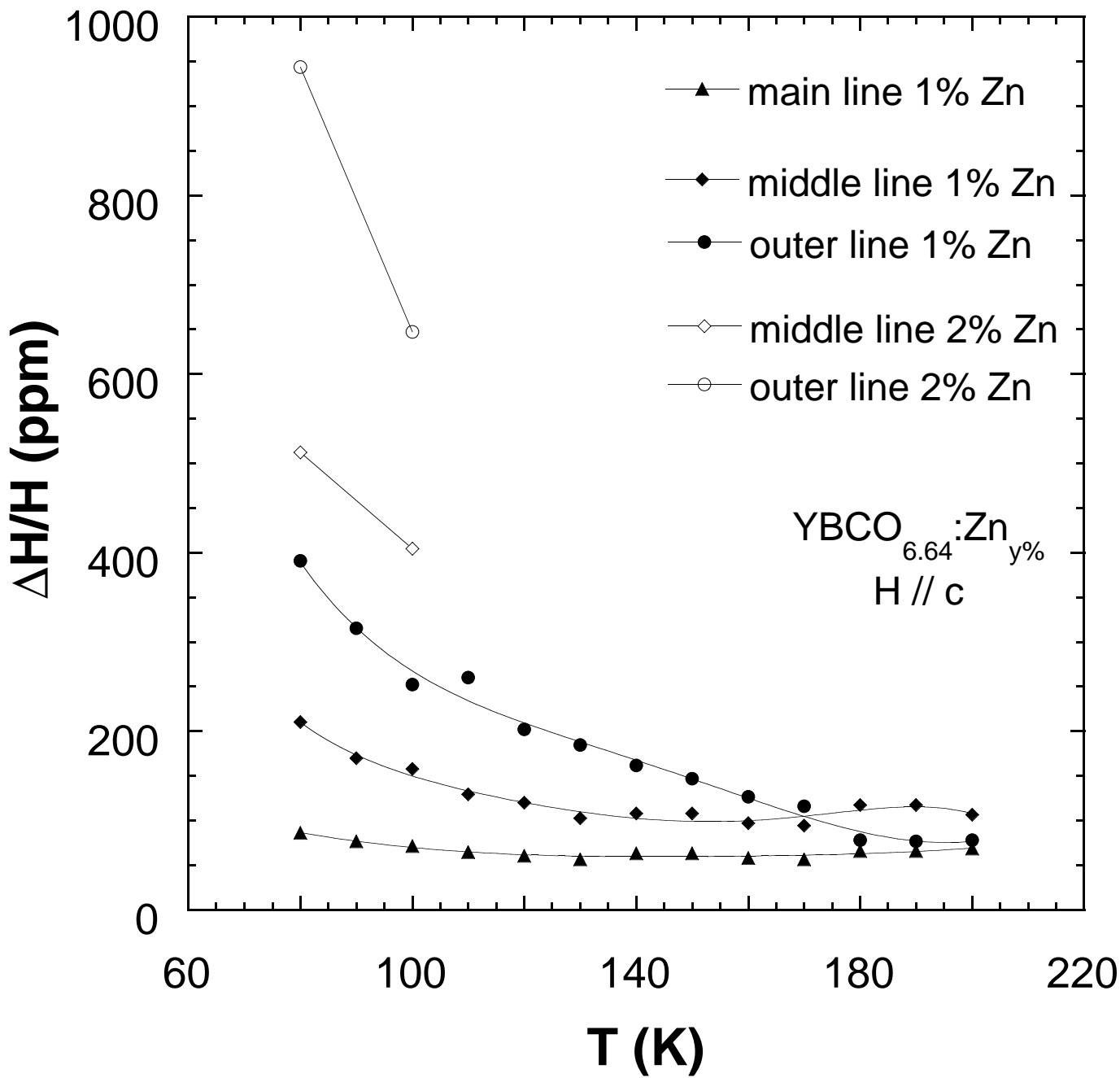
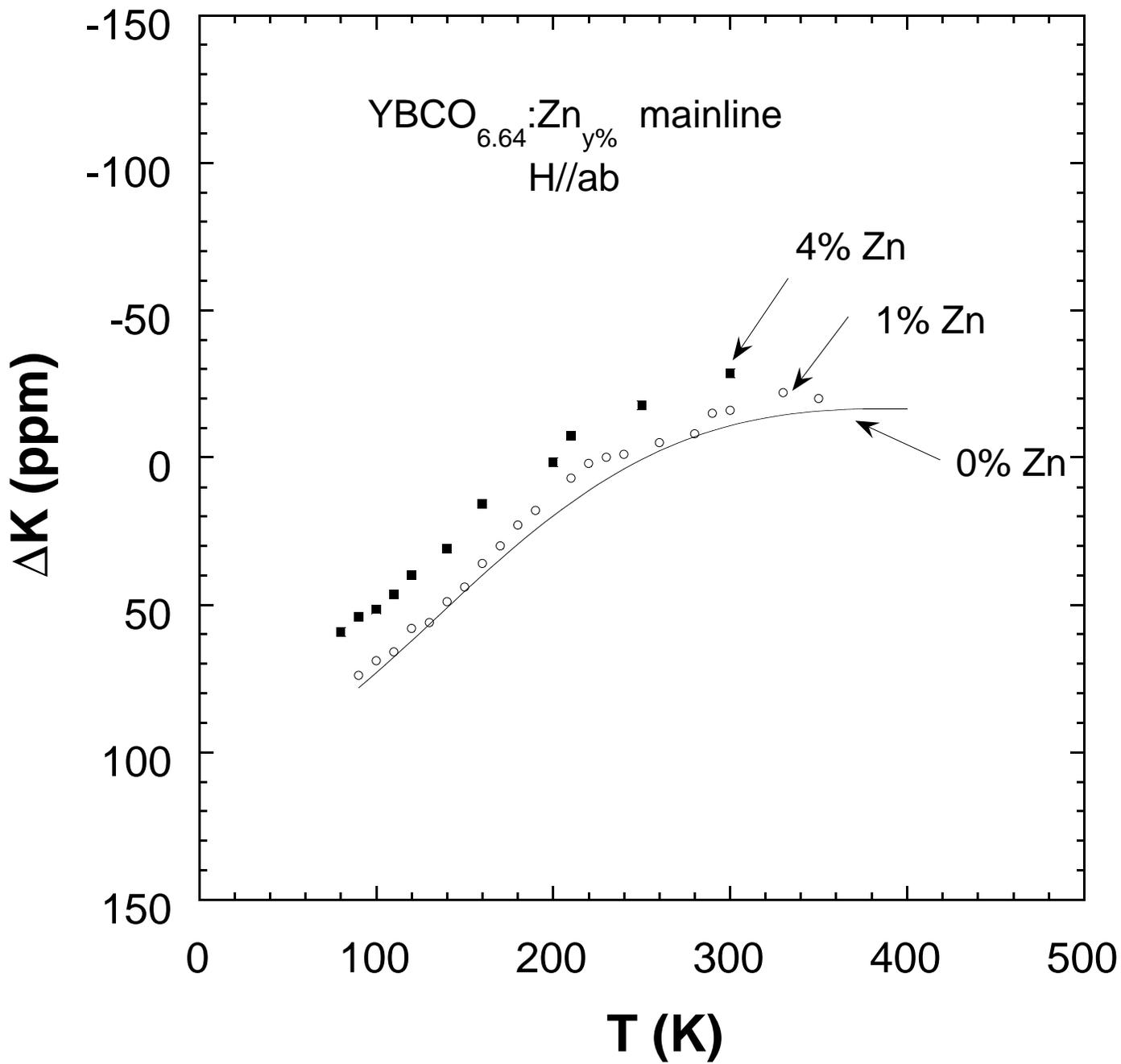
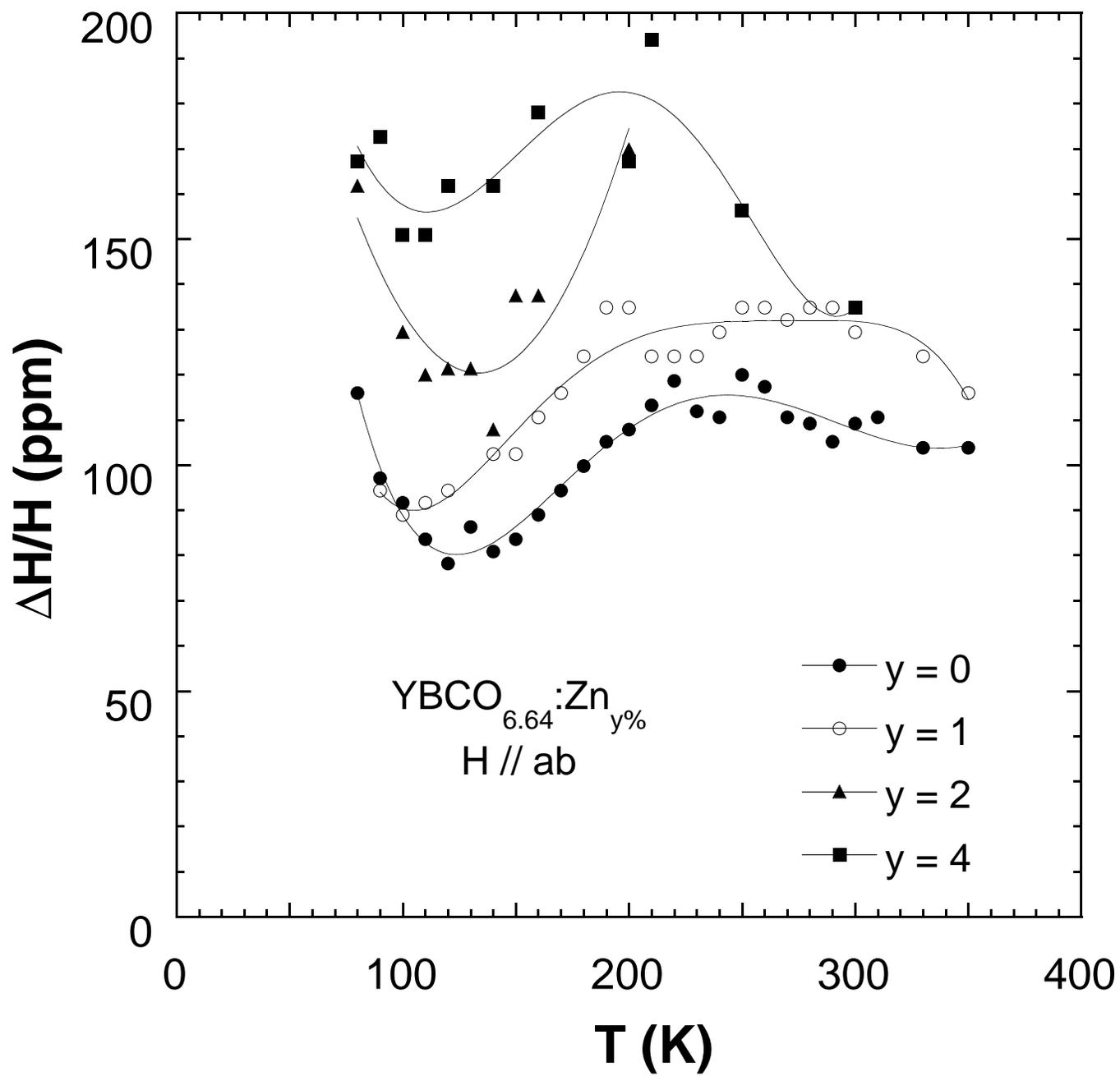
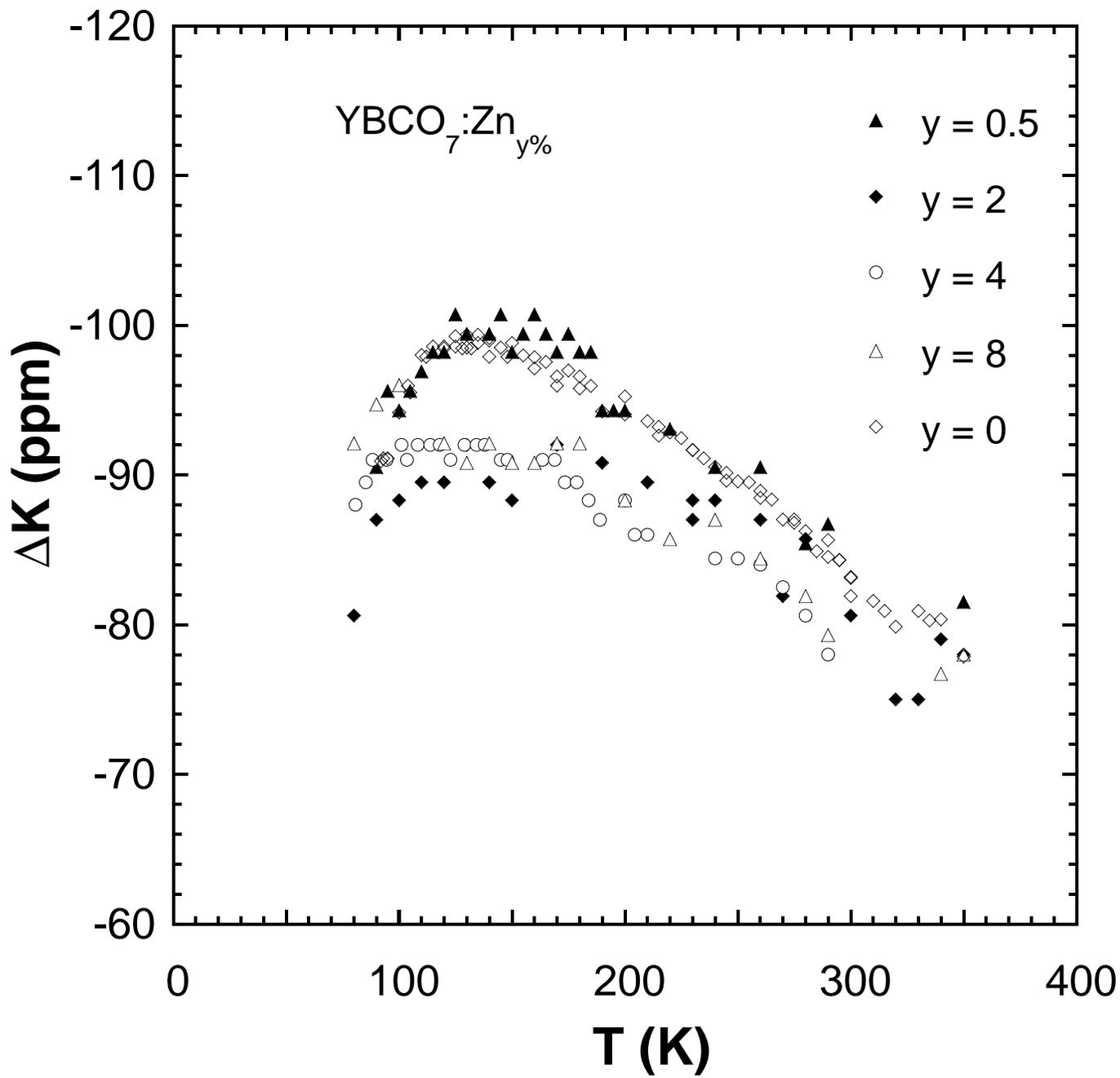
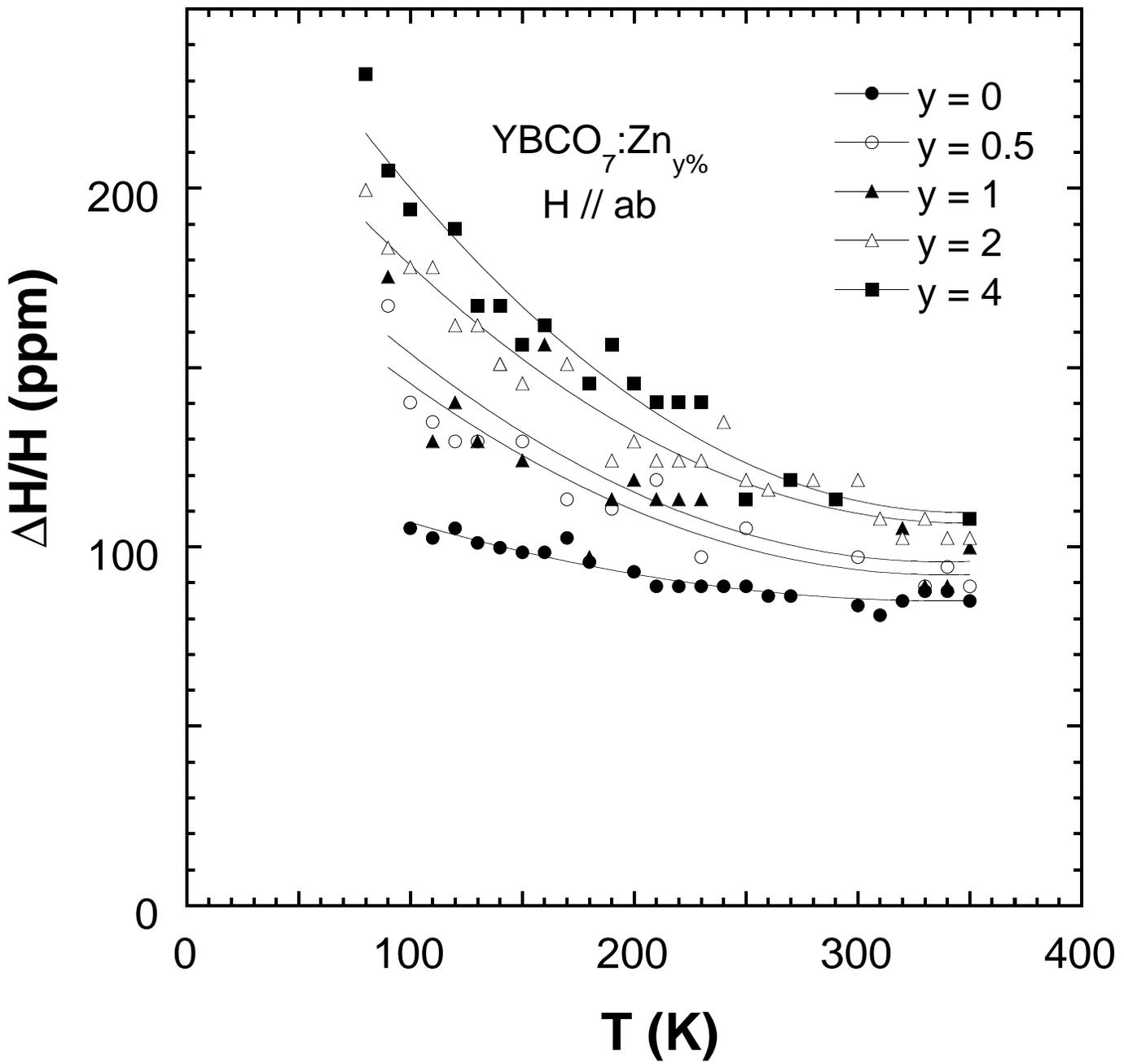
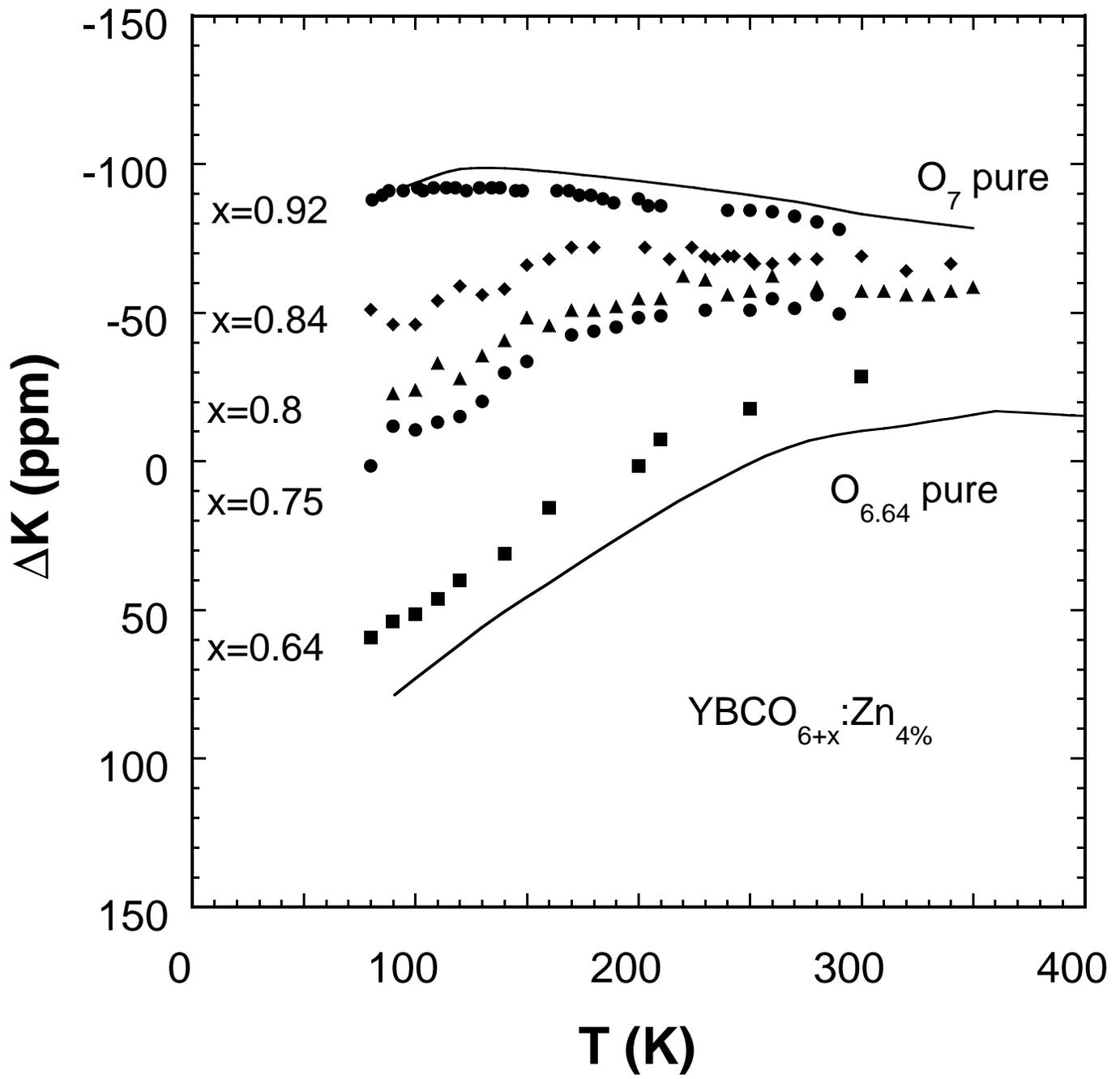
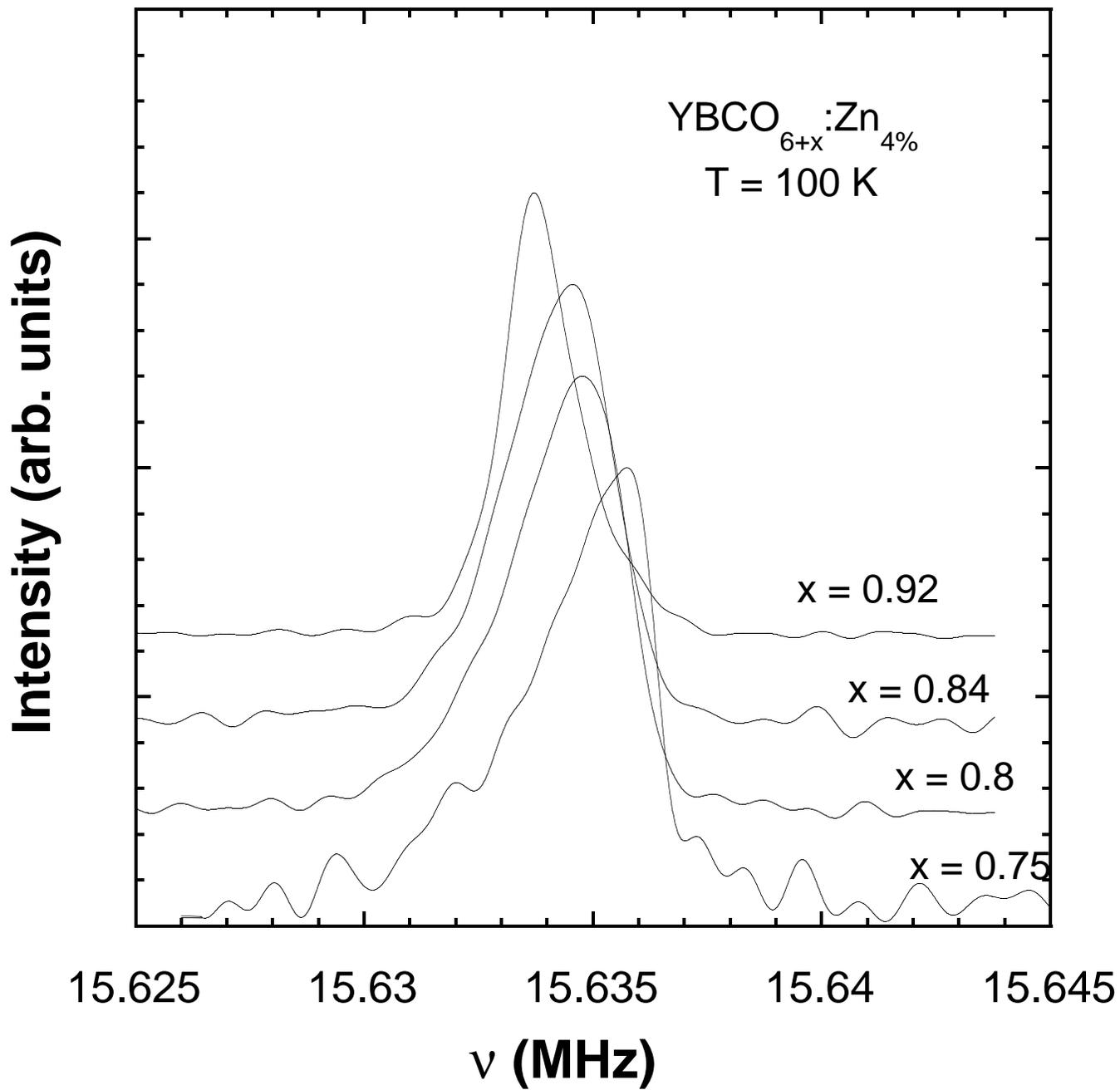
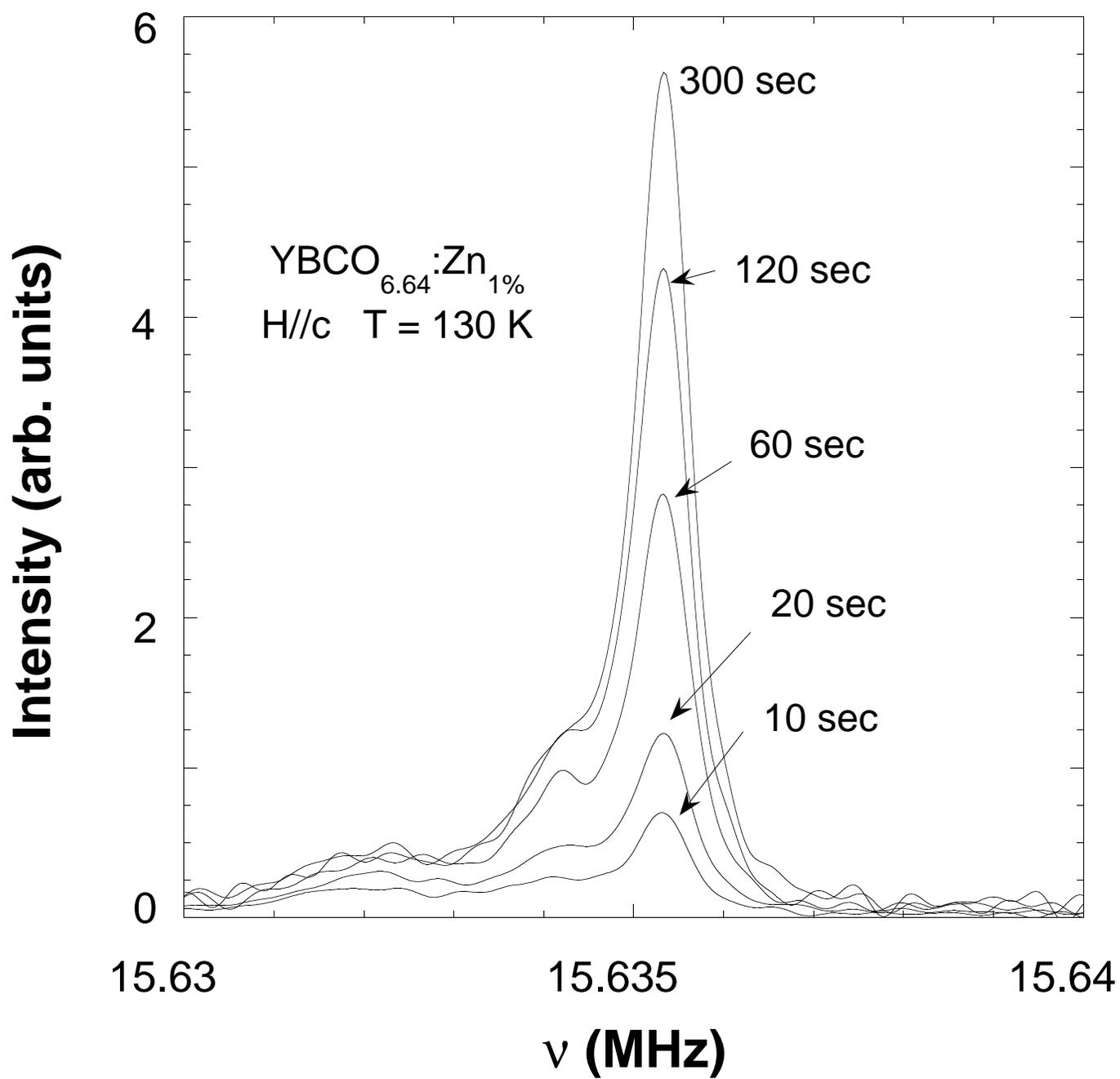
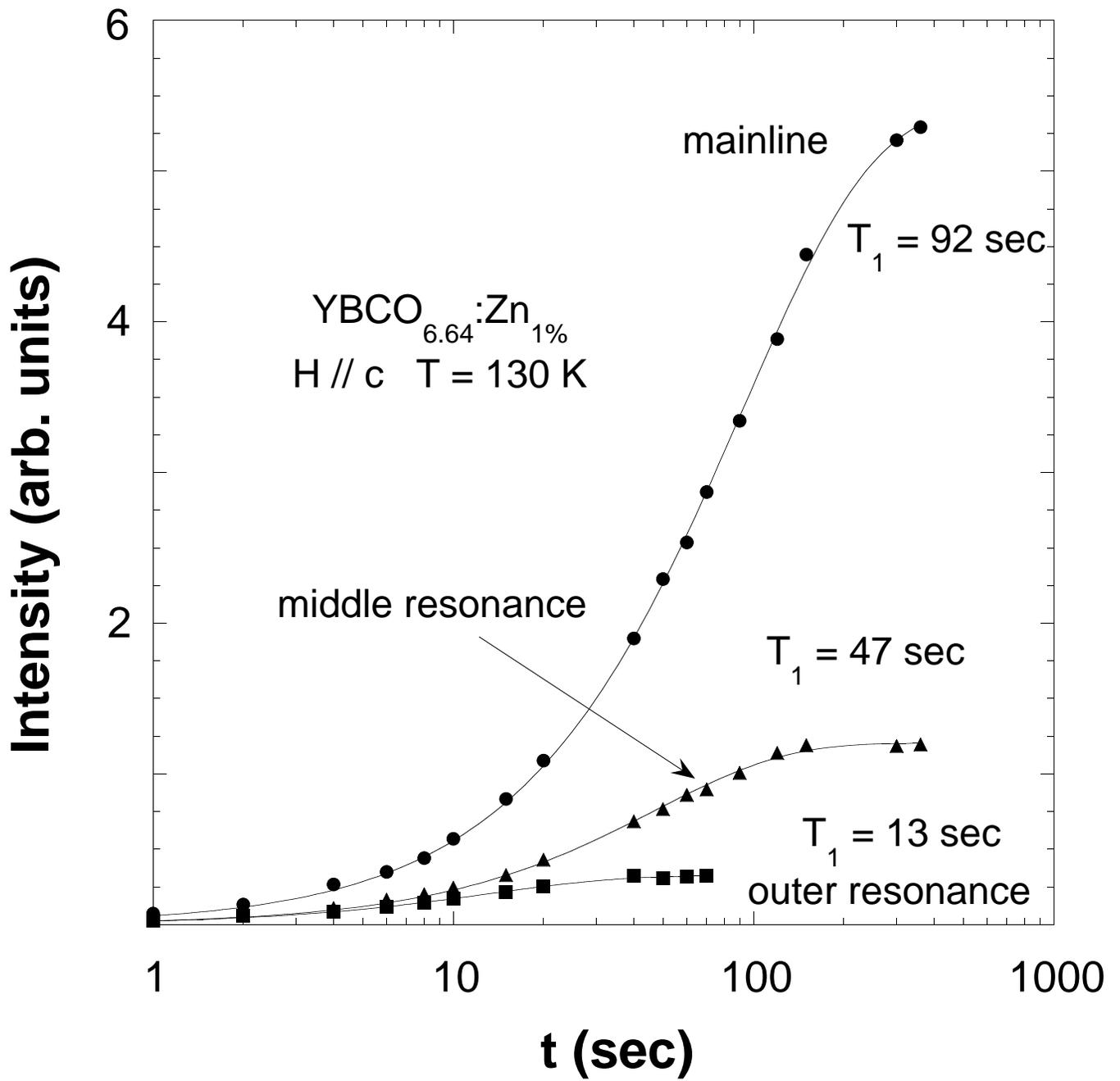
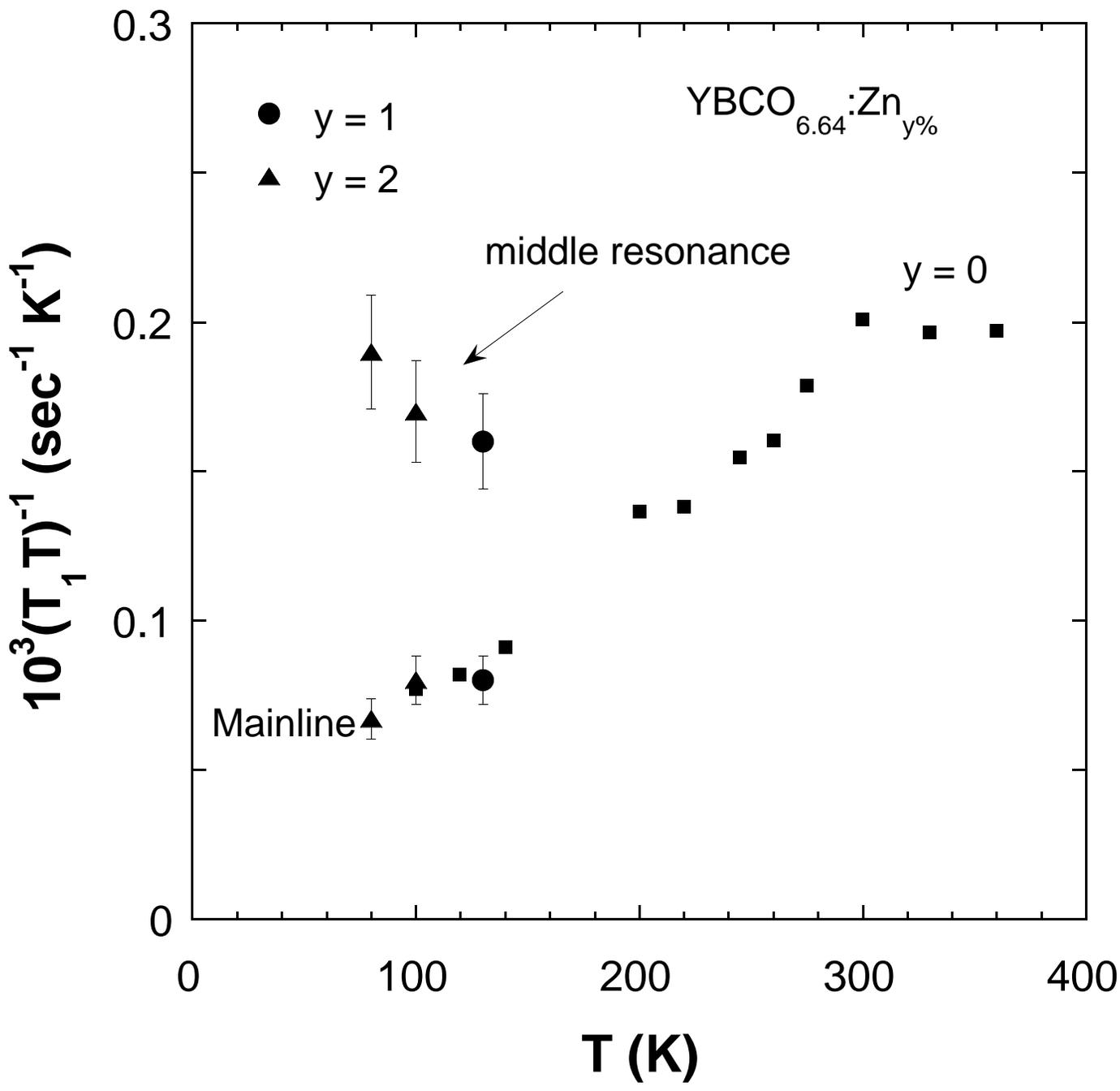
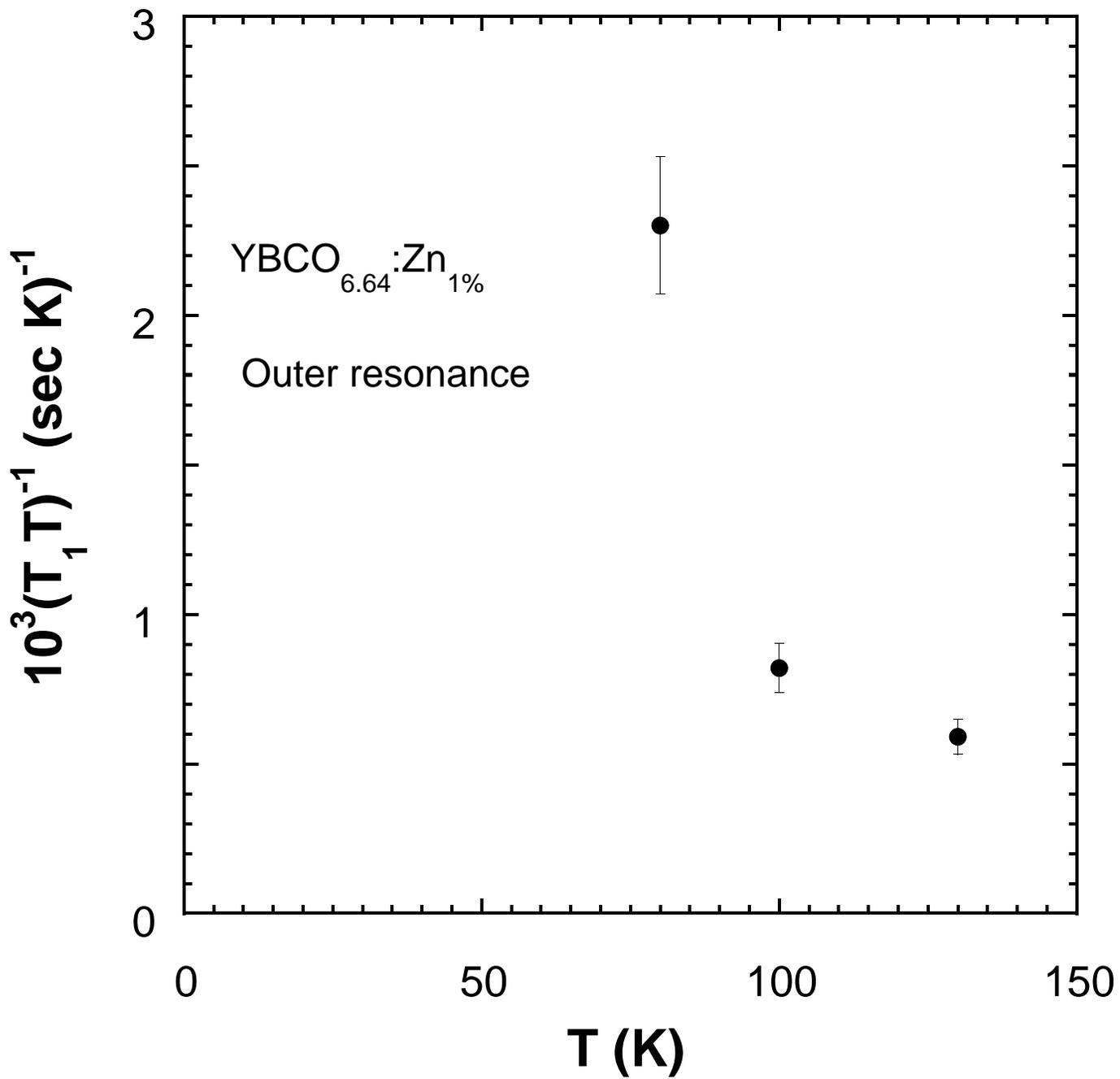
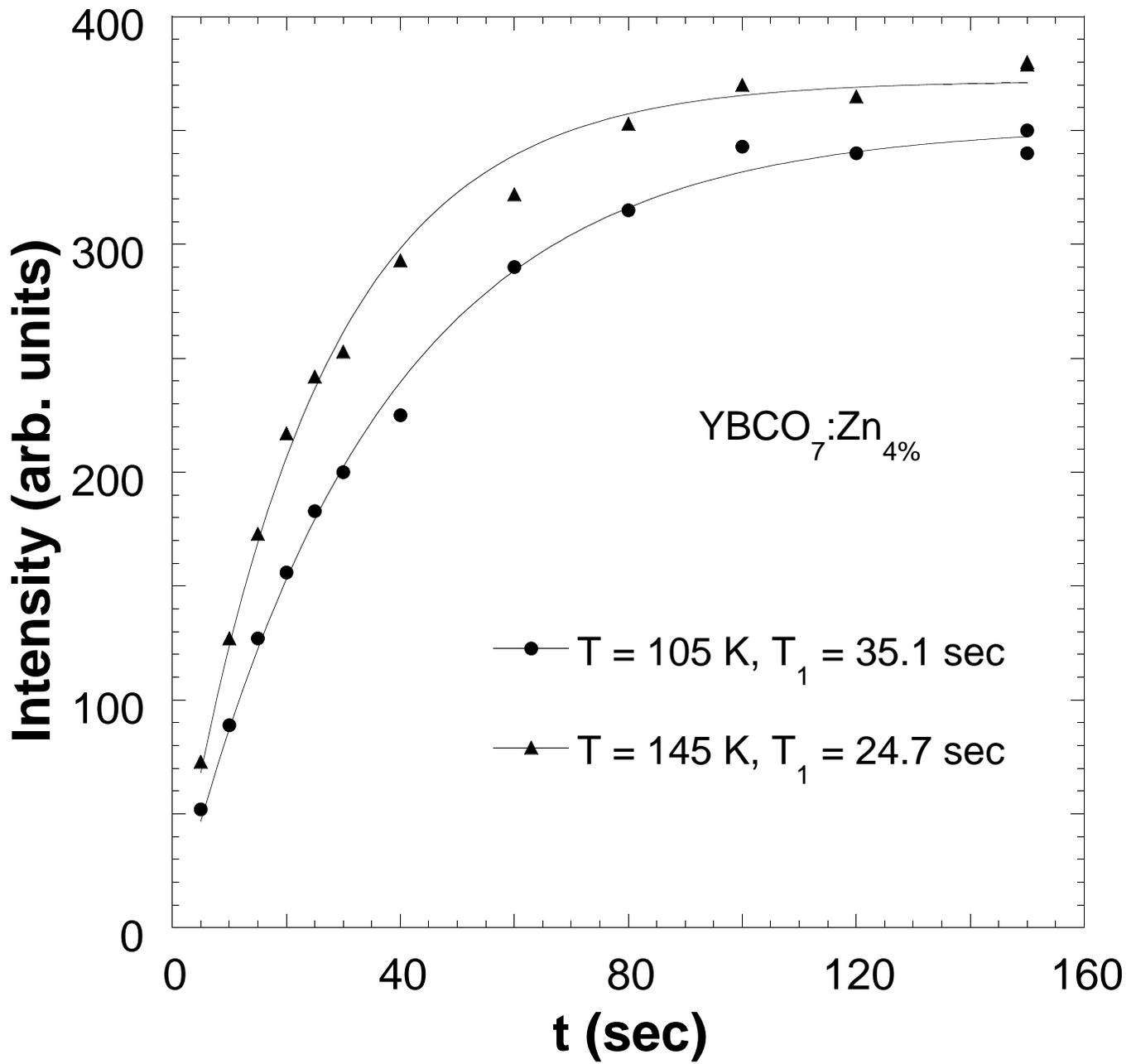
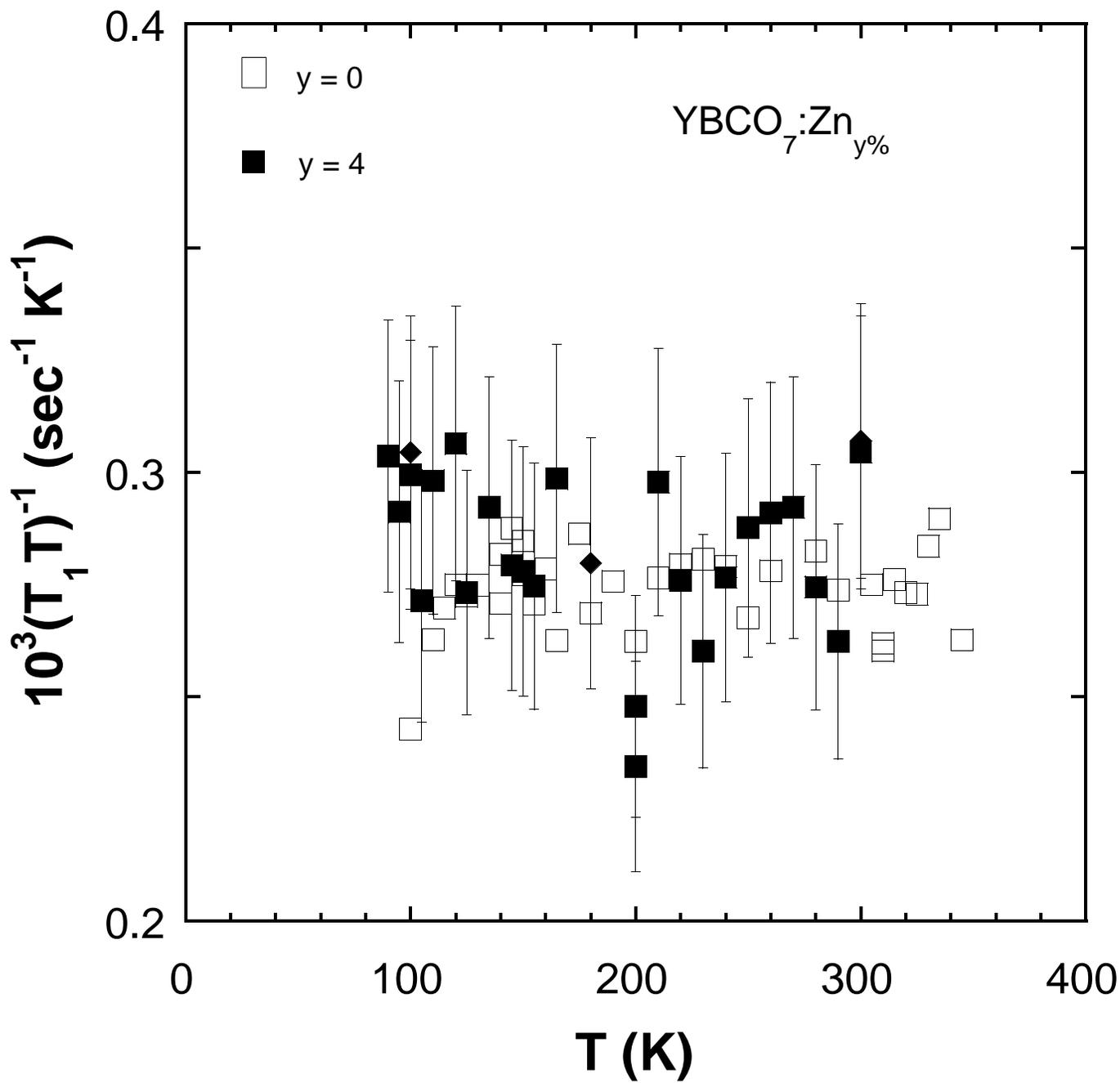
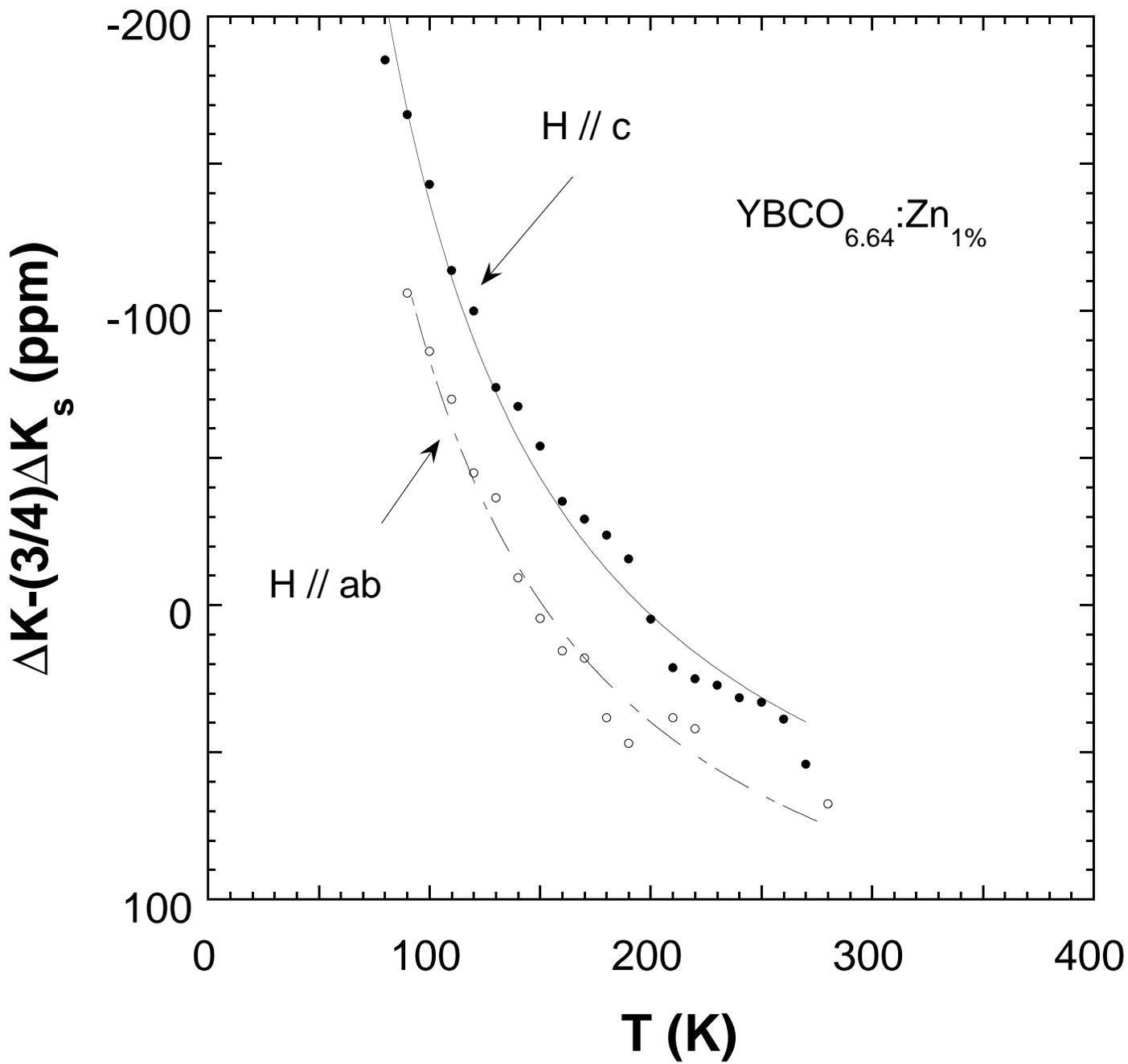
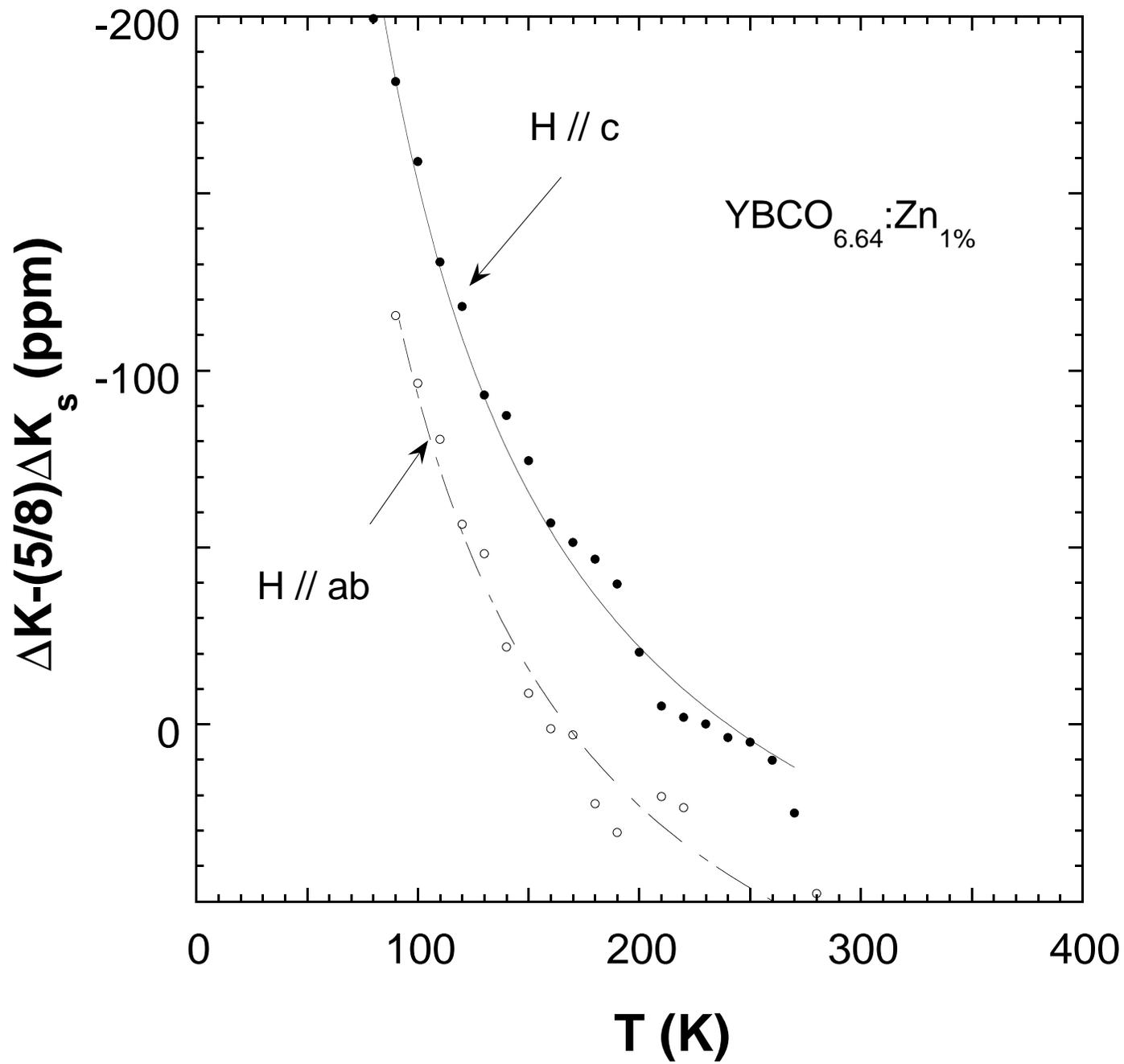
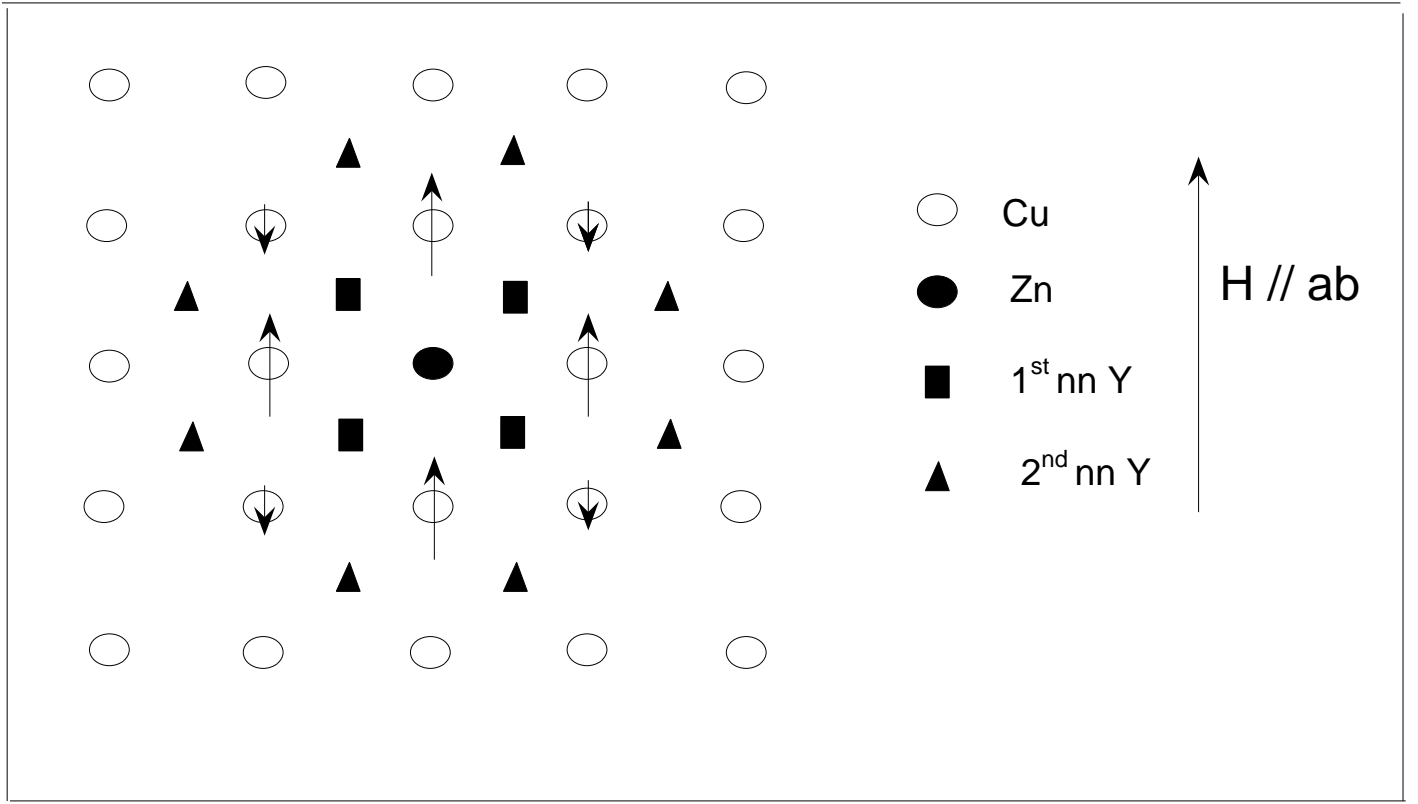
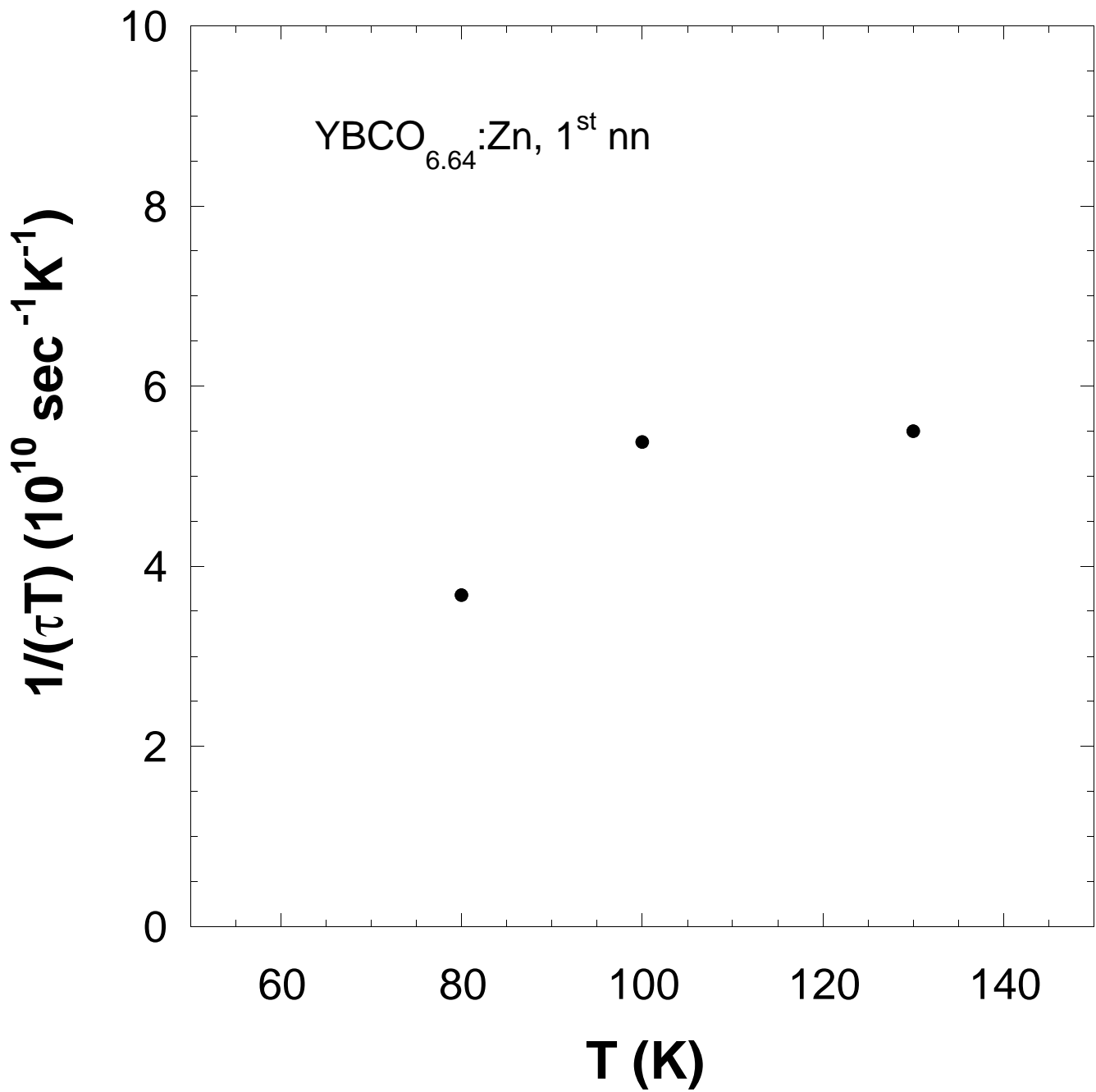
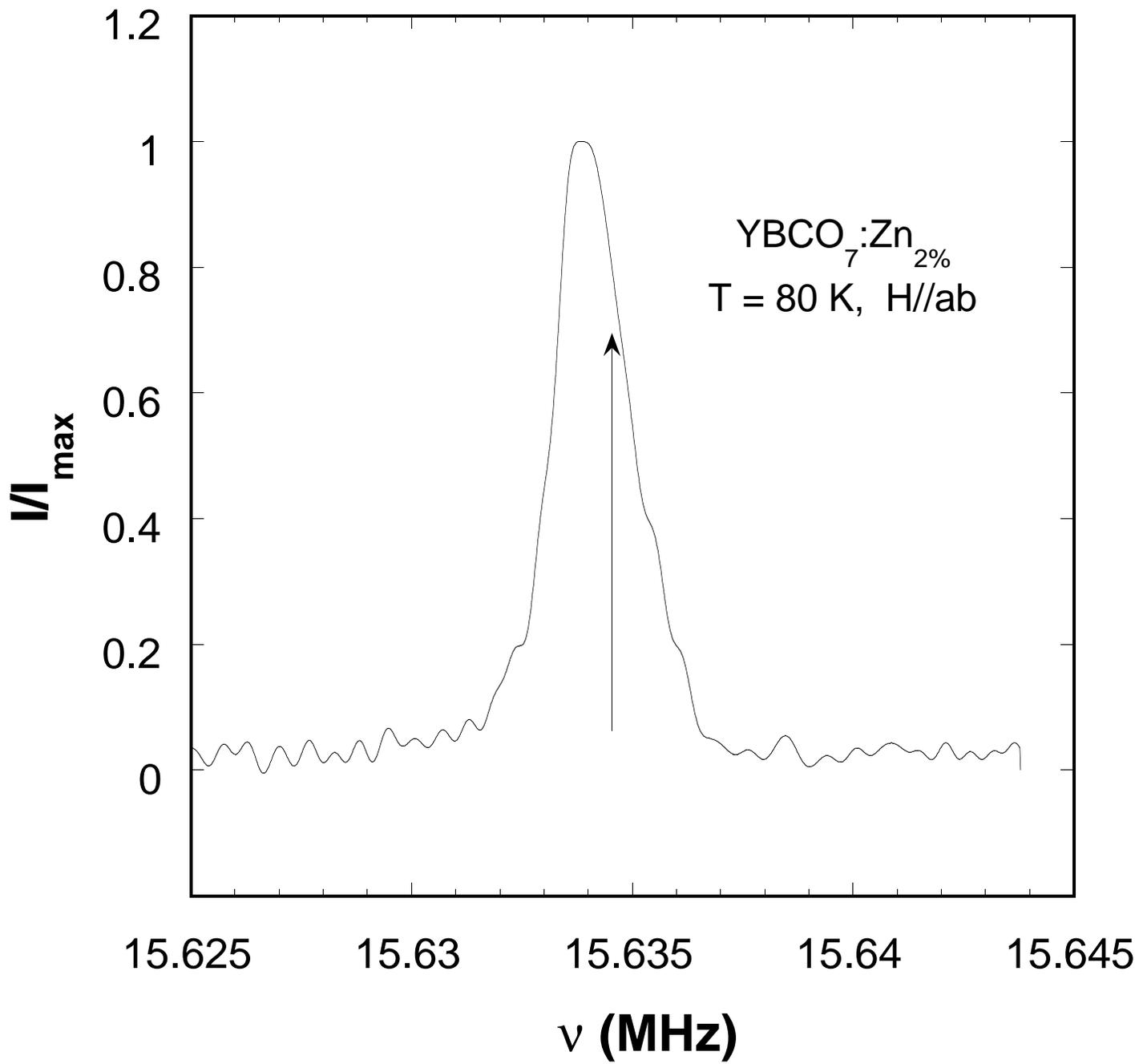
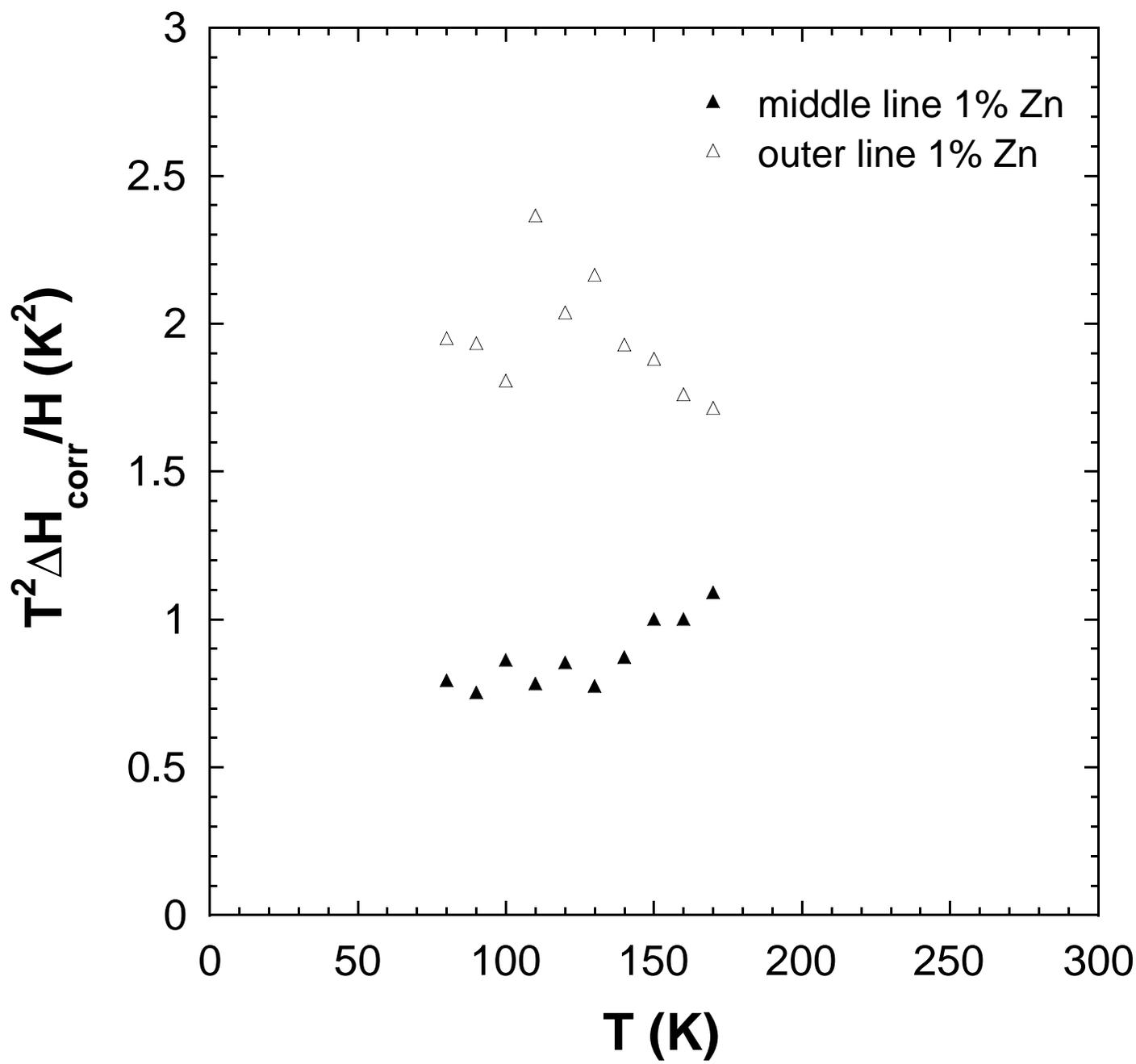
\end{document}